\newcount\Comments  
\documentclass[11pt]{article}
\pdfoutput=1
\usepackage[utf8x]{inputenc}
\usepackage{amssymb,amsmath,mathrsfs,enumerate,appendix}
\usepackage{graphicx,rotate,multicol, multirow}
\usepackage{cite}
\usepackage{braket}
\usepackage[normalem]{ulem}
\usepackage{wrapfig}
\usepackage{comment}
\usepackage[textfont=it,labelfont=bf]{caption}
\usepackage{bm}
\usepackage[colorlinks=true,
linkcolor=red,
urlcolor=magenta,
citecolor=blue]{hyperref}
\usepackage{lineno}
\usepackage{color}
\usepackage{bbold}
\usepackage{xcolor}
\long\def\rpl#1!!#2!!{\textcolor{red}{#1} \textcolor{blue}{#2}}

\def\@seccntformat#1{\@ifundefined{#1@cntformat}%
	{\csname the#1\endcsname\quad}
	{\csname #1@cntformat\endcsname}
}

\def\bar{\overline}

\makeatother

\newcommand{\bea}{\begin{eqnarray}}
\newcommand{\eea}{\end{eqnarray}}

\newcommand{\be}{\begin{equation}}
\newcommand{\ee}{\end{equation}}
\def\beq{\begin{equation}}
\def\eeq{\end{equation}}
\newcommand{\ba}{\begin{eqnarray}}
\newcommand{\ea}{\end{eqnarray}}
\def\ifmath#1{\relax\ifmmode #1\else $#1$\fi}

\def \order(#1){{\cal O} \left(#1 \right)}

\definecolor{darkgreen}{rgb}{0,0.5,0}
\definecolor{purple}{rgb}{1,0,1}
\newcommand{\kibitz}[2]{\ifnum\Comments=1\textcolor{#1}{#2}\fi}

\evensidemargin=\oddsidemargin
\def\Eqn#1{Eq.\ (\ref{#1})}
\def\Eqs#1#2{Eqs.\ (\ref{#1}) and (\ref{#2})}

\allowdisplaybreaks

\title{\Large\bf 
	Soft Symmetry Breaking as a Nonstandard Source of Mass: Phenomenological Insights from the Two-Higgs-Doublet Model 
}

\author{
	\sf 
	Dipankar Das$^{a,}$\footnote{d.das@iiti.ac.in},
	Miguel Levy$^{b,}$\footnote{miguelpissarra.levy@unibas.ch}, Shreya Pandey$^{a,}$\footnote{shreyavatspandey@gmail.com}, 
	Ipsita Saha$^{c,}$\footnote{ipsita@iitm.ac.in}, Agnivo Sarkar$^{d,}$\footnote{agnivosarkar@hri.res.in}
	\\[3mm]
    \small\em
	$^a$Indian Institute of Technology (Indore), Khandwa Road, Simrol,
	Indore 453 552, India \\
	\small\em
	$^b$ Department of Physics, University of Basel, Klingelbergstrasse 82, CH-4056 Basel, Switzerland \\ 
	\small\em
	$^c$Department of Physics, Indian Institute of Technology Madras, Chennai 600036, India \\
	\small\em
	$^d$Regional Centre for Accelerator-based Particle Physics, Harish-Chandra Research Institute, \\
		\small\em
	HBNI, Chhatnag Road, Jhunsi, Prayagraj 211019, India.
}

\date{}

\begin{document}
	
	
\maketitle
\renewcommand*{\thefootnote}{\arabic{footnote}}
\setcounter{footnote}{0}
	
\begin{abstract}
The soft-breaking parameter, $m_{12}^2$, frequently appearing in the 2HDM scalar
potential is much more remarkable than being just a nonstandard parameter that
helps make the BSM scalars super heavy. In fact, as we show through
explicit calculations, it should be treated as the direct but concise embodiment
of new non-electroweak spontaneous symmetry breaking effects at very high energy
scales, wherein lies its quiddities. Consequently, it is argued that $m_{12}^2$
and the electroweak VEV serve as two distinct sources for the nonstandard scalar
masses, which are completely unrelated to each other.
Such distinctions allow us to define parameters that conveniently
capture the fraction of the nonstandard scalar masses derived from the electroweak
VEV. Finally, we demonstrate that constraints can already be
placed on such fractions from the current measurements of the diphoton signal
strength and from direct searches of new nonstandard scalar resonances in the
diphoton channel.
\end{abstract}
	
	\maketitle
	
	\section{Introduction}
The two Higgs-doublet model (2HDM)\cite{Branco:2011iw} provides a minimal yet rich extension of the Standard Model (SM) scalar sector 
by introducing a second complex $SU(2)_L$ doublet with hypercharge $Y = +\frac{1}{2}$ ($Q = T_3 + Y$). In the version
where CP-symmetry is assumed to be respected by the scalar potential, the spectrum of physical scalars can be categorized
into two CP-even scalars (one of which should be identified with the SM-like Higgs boson that has been found at the LHC),
one pseudo-scalar and a pair of charged scalars. One of the usual features in the 2HDM formulations is the  imposition of
a discrete symmetry -- typically a $Z_2$-symmetry -- to prevent tree-level flavor-changing neutral-currents (FCNCs).
However, exact symmetries can be overly restrictive, especially when the decoupling of non-standard scalars is concerned~\cite{Bhattacharyya:2014oka,Faro:2020qyp,Carrolo:2021euy}. To address this,
soft-breaking terms are usually introduced in the scalar potential. The soft-breaking parameter, often denoted as $m_{12}^2$, plays a crucial role in 
making the nonstandard scalars decouple, especially when the discrete $Z_2$-symmetry in the scalar potential is spontaneously broken.
Understanding the origin of this soft-breaking term is essential, as it can offer insights into the ultraviolet (UV) completions of the model
and hint at deeper symmetries or mechanisms at higher energy scales. 
We study these aspects here within the framework of the 2HDM and highlight the structural differences between softly-broken discrete and continuous symmetries when requiring the complete theory to remain fully symmetric. In particular, we explicitly demonstrate how the soft-breaking parameter can be directly related to a nonstandard mass scale that is independent of the electroweak vacuum expectation value (VEV).

In this work, thus,  we aim to demystify the origin and implications of the soft-breaking term in the 2HDM scalar potential by analyzing its role in shaping the physical scalar spectrum. To this end, we systematically decompose the mass expressions of the nonstandard scalars into two distinct types of contributions: those directly tied to the electroweak VEV, and those that arise from independent mass parameters, particularly those associated with soft symmetry breaking. This decomposition serves to disentangle the effects that are inherently linked to electroweak symmetry breaking from those that encode physics at scales not directly related to the Higgs mechanism
at the electroweak scale.

To facilitate this analysis, we introduce a convenient parametrization that allows for a clear phenomenological separation between electroweak VEV-induced and non-electroweak contributions to the nonstandard scalar masses. This parametrization is especially useful in the decoupling regime, where the SM-like Higgs boson exhibits behavior that closely resembles the
corresponding SM expectations.

More importantly, our framework provides a pathway to probe the electroweak contributions to the nonstandard scalar masses through collider observables, particularly in the TeV-scale energy domain. By correlating theoretical mass structures with experimental data, we offer means to constrain or even measure the extent to which the electroweak VEV influences the heavier scalar states. Such connections are crucial, especially in models where the soft-breaking parameter is not an arbitrary input, but instead emerges from deeper UV dynamics or symmetry considerations.
In this context, we highlight the role of neutral nonstandard Higgs searches, particularly in the di-photon final state, as a potential probe of the electroweak-induced mass contributions. 
In particular, it allows us to test the extent to which electroweak symmetry breaking governs the masses of the nonstandard scalars, thereby offering indirect insights into the very nature of spontaneous symmetry breaking and its potential interplay with softly-broken discrete symmetries.
%

The article is organized as follows. In Sec.~\ref{sec:Z2-2HDM}, we provide an overview of the softly-broken $Z_2$-symmetric 2HDM as an illustrative example, including an explicit calculation of the scalar potential and mass diagonalization. In Sec.~\ref{sec:Z2-2HDMS}, we extend the discussion to a high-scale $Z_2$-symmetric theory with a singlet-extended 2HDM scenario, demonstrating how the soft-breaking parameter in the low-energy 2HDM emerges from the spontaneous breaking of the high-scale symmetry.
To highlight constructional differences, Sec.~\ref{sec:U1-2HDM} considers a softly-broken $U(1)$-symmetric theory and shows that the corresponding UV-complete realization is a $Z_3$-symmetric 2HDM with an additional complex singlet scalar. In Sec.~\ref{s:param}, we introduce a parametrization for the electroweak contributions to the nonstandard scalar masses, which is then employed in the subsequent phenomenological analyses. Theoretical constraints are discussed in Sec.~\ref{sec:theorybound}, and the implications of Higgs to diphoton searches are presented in Sec.~\ref{s:diphoton}.
Our combined results, incorporating both theoretical and direct search constraints from the LHC, are presented in Sec.~\ref{s:results}. Finally, Sec.~\ref{s:summary} summarizes our findings and concludes.

\section{The two-Higgs-doublet model: a case study}
\label{sec:Z2-2HDM}
%
 In this section, we begin by revisiting the scalar sector of a 2HDM featuring a softly-broken $Z_2$-symmetry. This initial discussion aims to provide a preliminary adumbration of the key concepts to help orient the reader before we explore them in greater detail in subsequent sections.
The most general 2HDM scalar potential with a softly-broken $Z_2$-symmetry
is given by~\cite{Branco:2011iw,Bhattacharyya:2015nca},
\begin{eqnarray}
	V(\Phi_{1}, \Phi_{2}) & = & m^{2}_{11}\Phi^{\dagger}_{1}\Phi_{1} +  m^{2}_{22}\Phi^{\dagger}_{2}\Phi_{2} - \left[m^{2}_{12}\Phi^{\dagger}_{1}\Phi_{2} + \text{h.c.}\right] + \frac{\lambda_{1}}{2}\left(\Phi^{\dagger}_{1}\Phi_{1}\right)^{2} + \frac{\lambda_{2}}{2}\left(\Phi^{\dagger}_{2}\Phi_{2}\right)^{2}  \nonumber \\ 
	&& + \lambda_{3}\left(\Phi^{\dagger}_{1}\Phi_{1}\right)\left(\Phi^{\dagger}_{2}\Phi_{2}\right) + \lambda_{4}\left(\Phi^{\dagger}_{1}\Phi_{2}\right)\left(\Phi^{\dagger}_{2}\Phi_{1}\right) + \left[\frac{\lambda_{5}}{2}\left(\Phi^{\dagger}_{1}\Phi_{2}\right)^{2} + \text{h.c.}\right] \,.
	\label{e:2hdm_pot}     
\end{eqnarray}
We will be working under the assumption of CP-conservation in the scalar sector.
The bilinear term proportional to  $m^{2}_{12}$ encodes the effects of soft $Z_2$ breaking. It should be noted that in the limit $\lambda_5=0$, the symmetry of the
scalar potential will be elevated to a softly-broken $U(1)$ symmetry. 
We expand the $SU(2)_L$ Higgs-doublets in terms of their component fields as follows:
\begin{eqnarray}
	\label{e:field_definition}
	\Phi_j = \begin{pmatrix}
		w_j^+ \\
		(v_j+h_j+ i z_j)/\sqrt{2}
	\end{pmatrix} \,, \qquad (j=1,2) \,.
\end{eqnarray}
 After the
spontaneous symmetry breaking~(SSB), the total electroweak VEV $v$, can be
identified in terms of the individual VEVs of the doublets $v_j$,
\begin{eqnarray}
	v^2 = v_1^2 + v_2^2 =  (246 \,\, {\rm GeV})^2 \,,
	\qquad {\rm with,}~~ \tan \beta = \frac{v_2}{v_1}\,.
	\label{e:2hdm_vev}
\end{eqnarray}
%
 The minimization conditions can be used to 
trade the bilinears $m_{11}^2$ and $m_{22}^2$ in terms of the VEVs $v_1$ and $v_2$
as follows: 
\begin{subequations}
	\begin{eqnarray}
		m_{11}^2 &=&  m_{12}^2 \frac{v_2}{v_1} - \frac{\lambda_1}{2} v_1^2 
		- \frac{1}{2}\left
		(\lambda_3+\lambda_4+\lambda_5\right)v_2^2 \,, \\
		m_{22}^2 &=&  m_{12}^2 \frac{v_1}{v_2} - \frac{\lambda_2}{2} v_2^2  -\frac{1}{2}\left
		(\lambda_3+\lambda_4+\lambda_5\right)v_1^2  \,.
	\end{eqnarray}
	\label{e:minim_2hdm}
\end{subequations}
At this point we note that the scalar potential of \Eqn{e:2hdm_pot}
contains eight parameters including the bilinear parameters
$m_{11}^2$, $m_{22}^2$ which we have  traded for the two
VEVs, $v_1$ and $v_2$  or equivalently $v$ and $\tan\beta$.
The five quartic couplings can be
exchanged for four physical masses (two CP-even scalars,
one CP-odd scalars and a pair of charged scalars) and the neutral mixing angle,
namely $\alpha$, all of which will be defined in the following subsection. 
%

 \subsection{Physical eigenstates}
Let us first discuss in detail the mass matrices that arise from the scalar
potential and the resulting physical mass eigenstates.
Since all the parameters in the scalar
 potential are assumed to be real, there will be no mass mixing between the $h_j$
 and the $z_j$ fields. Therefore, the neutral mass eigenstates can be conveniently
 categorized as CP-even and CP-odd scalars.
In the pseudoscalar sector, the corresponding mass matrix is derived as shown below, where the unphysical fields are specified in Eq.~\eqref{e:field_definition}.
 \begin{subequations}
 	\begin{eqnarray}
 		V^{\rm mass}_P &=& \begin{pmatrix}
 			z_1 & z_2 
 		\end{pmatrix} \, \frac{{\cal M}_P^2}{2} \, \begin{pmatrix}
 			z_1\\  z_2\\
 		\end{pmatrix} \,, \\
{\rm with,} \qquad {\cal M}_P^2 &=& \begin{pmatrix}
 			m_{12}^2 \frac{v_2}{v_1} - \lambda_5 v_2^2   & -m_{12}^2 + \lambda_5 v_1 v_2 \\
 			-m_{12}^2 + \lambda_5 v_1 v_2 &	m_{12}^2\frac{ v_1}{v_2} - \lambda_5 v_1^2 \\
 		\end{pmatrix} \,, \\ 
 		&=& \left[\frac{m_{12}^2}{v_1 v_2} -\lambda_5 \right] \begin{pmatrix} v_2^2 & -v_1v_2 \\
 			-v_1 v_2 & v_1^2 \\
 		\end{pmatrix} \,. \label{e:mpsq}
 	\end{eqnarray}
  	\label{e:mps_2hdm}
 \end{subequations}
Similarly, the charged scalar mass matrix can be written as,
\begin{subequations}
	\begin{eqnarray}
		V^{\rm mass}_C &=& \begin{pmatrix}
			w_1^+ & w_2^+ 
		\end{pmatrix} \, {\cal M}_C^2 \, \begin{pmatrix}
			w_1^-\\  w_2^-\\
		\end{pmatrix} \,, \\
{\rm with,} \qquad		{\cal M}_C^2 &=& \begin{pmatrix}
			m_{12}^2 \frac{v_2}{v_1} -\frac{1}{2} \left(\lambda_4+\lambda_5\right) v_2^2  &	-m_{12}^2 +\frac{1}{2} \left(\lambda_4+\lambda_5\right) v_1 v_2 \\
			-m_{12}^2 +\frac{1}{2} \left(\lambda_4+\lambda_5\right) v_1 v_2 &	m_{12}^2\frac{ v_1}{v_2} -\frac{1}{2} \left(\lambda_4+\lambda_5\right)v_1^2 \\
		\end{pmatrix} \,, \\ 
		&=& \left[\frac{m_{12}^2}{v_1 v_2} -\frac{1}{2} \left(\lambda_4+\lambda_5\right)\right] \begin{pmatrix} v_2^2 & -v_1v_2 \\
			-v_1 v_2 & v_1^2 \label{e:mc_2hdm}\\
		\end{pmatrix} \,. \label{e:mcsq}
	\end{eqnarray}
	\label{e:mcs_2hdm}
\end{subequations}
The forms of the matrices in \Eqs{e:mpsq}{e:mcsq} suggest that 
${\cal M}_C^2$ and ${\cal M}_P^2$ may be diagonalized through an orthogonal 
transformation involving the matrix ${\cal O}_\beta$, defined as
 \begin{eqnarray}
 	{\cal O}_\beta =  \frac{1}{v}\begin{pmatrix}
 		v_1 & v_2 \\
 		-v_2 & v_1 \,
 	\end{pmatrix} \equiv \begin{pmatrix}
 		\cos \beta & \sin \beta \\
 		-\sin \beta & \cos \beta \,
 		\end{pmatrix} \,. 	
 	\label{e:Obeta}
 \end{eqnarray}
The fields in the physical basis are then obtained as
\begin{eqnarray}
	\begin{pmatrix}	G^0 \\  A \end{pmatrix}
	= {\cal O}_\beta \begin{pmatrix} z_1 \\  z_2 \end{pmatrix},
	\quad
	\begin{pmatrix}	G^\pm \\  H^\pm \end{pmatrix}
	= {\cal O}_\beta \begin{pmatrix} w_1^\pm \\  w_2^\pm \end{pmatrix}, 
\end{eqnarray}
where, $G^0$ and $G^\pm$ stand for the neutral and the charged Goldstones
respectively. The fields $A$ and $H^\pm$ denote the physical pseudoscalar and the
charged scalars with masses $m_A$ and $m_C$ respectively.
The diagonal forms of ${\cal M}_C^2$ and ${\cal M}_P^2$ can then be written as
\begin{eqnarray}
{\cal O}_\beta\cdot {\cal M}_P^2 \cdot {\cal O}_\beta^T = 
\begin{pmatrix}	0 & 0 \\ 0 & m_A^2 \end{pmatrix},
\quad
{\cal O}_\beta\cdot {\cal M}_C^2 \cdot {\cal O}_\beta^T = 
\begin{pmatrix}	0 & 0 \\ 0 & m_C^2 \end{pmatrix}.
\label{e:mcmp_diag_2hdm}
\end{eqnarray}
The explicit expressions for the mass-squared eigenvalues are given by
\begin{subequations}
 \begin{eqnarray}
 	m_{C}^2 &=& \frac{m_{12}^2}{\sin\beta\cos\beta} - \frac{v^2}{2} (\lambda_4+\lambda_5) \equiv \Lambda^2 - \frac{v^2}{2} (\lambda_4+\lambda_5) \\
 	m_{A}^2 &=& \frac{m_{12}^2}{\sin\beta \cos\beta} - \lambda_5 v^2 
 	\equiv \Lambda^2 - \lambda_5 v^2 \,,
 \end{eqnarray}
\label{e:ma_mc_2hdm}
\end{subequations}
where we have introduced a very suggestive shorthand for the soft $Z_2$-breaking
parameter as follows:
\begin{eqnarray}
{\Lambda^2} = \frac{m_{12}^2}{\sin\beta \cos \beta} \,.
\label{e:Lamdef}
\end{eqnarray}
%
For the neutral CP-even scalar sector, we find the mass matrix as,
 \begin{subequations}
	\begin{eqnarray}
		V^{\rm mass}_S &=& \begin{pmatrix}
			h_1 & h_2 
		\end{pmatrix} \, \frac{{\cal M}_S^2}{2} \, \begin{pmatrix}
			h_1\\ h_2\\
		\end{pmatrix} \,, \\
{\rm with,} \qquad	{{\cal M}_S^2} &=& \frac{\Lambda^2}{v^2} \begin{pmatrix}
	v_2^2  & -v_1 v_2 \\
	-v_1 v_2 &	v_1^2 \end{pmatrix}
	+ \begin{pmatrix}
	\lambda_1 v_1^2  & (\lambda_3+\lambda_4+\lambda_5) v_1 v_2 \\
	(\lambda_3+\lambda_4+\lambda_5) v_1 v_2 &	\lambda_2 v_2^2 \end{pmatrix} \,.
	\label{e:MS2}
\end{eqnarray}
\label{e:cpeven_2hdm}
\end{subequations}
This mass squared matrix in the CP-even sector is assumed to be diagonalized
via the following rotation:
\begin{eqnarray}
	\begin{pmatrix}
		h \\ H
	\end{pmatrix} = \begin{pmatrix} \cos\alpha & \sin\alpha \\
	-\sin\alpha & \cos\alpha 
	\end{pmatrix} \begin{pmatrix} h_1 \\ h_2
	\end{pmatrix} \equiv {\cal O}_\alpha \begin{pmatrix} h_1 \\ h_2
	\end{pmatrix}.
 \label{e:h125_2hdm}
\end{eqnarray}
which requires, 
\begin{eqnarray}
	{\cal O}_\alpha\cdot {\cal M}_S^2 \cdot {\cal O}_\alpha^T = 
	\begin{pmatrix}	m_h^2 & 0 \\ 0 & m_H^2 \end{pmatrix}\,.
\end{eqnarray}
It will thus entail the following relations
\begin{subequations}
\begin{eqnarray}
	m_h^2 \cos^2 \alpha + m_H^2 \sin^2 \alpha &=& \Lambda^2\sin^2\beta
	+\lambda_1 v^2\cos^2\beta \,, \\
	 (m_H^2 - m_h^2) \cos\alpha \sin\alpha &=& \left\{ \Lambda^2 - (\lambda_3+\lambda_4+\lambda_5) v^2\right\} \cos\beta \sin\beta \,,  \\
	m_h^2 \sin^2 \alpha + m_H^2 \cos^2 \alpha &=& \Lambda^2\cos^2\beta
	+\lambda_2 v^2\sin^2\beta \,.
\end{eqnarray}
\label{e:rotalpha}
\end{subequations}
Inverting \Eqs{e:ma_mc_2hdm}{e:rotalpha}, we can express the quartic parameters in the
2HDM scalar potential as follows: 
\begin{subequations}
	\begin{eqnarray}
		\lambda_1 &=& 	\frac{1}{v^2 \cos^2\beta} \left(m_h^2 \cos^2\alpha + 
		m_H^2 \sin^2 \alpha\right) - \frac{\Lambda^2}{v^2}  \tan^2\beta \,, \label{e:lam1_2hdm}\\
		\lambda_2 &=& \frac{1}{v^2 \sin^2 \beta} \left(m_H^2 \cos^2 \alpha +  
		m_h^2 \sin^2 \alpha\right) - \frac{\Lambda^2}{v^2}  \frac{1}{\tan^2 \beta} \,, \label{e:lam2_2hdm}\\
		\lambda_3 &=& \frac{1}{v^2} \left(2 m_C^2 - \Lambda^2 - \left(m_H^2 - m_h^2\right) \frac{
		\cos \alpha  \sin \alpha}{\cos \beta \sin \beta}\right)\,, \label{e:lam3_2hdm}\\ 
		\lambda_4 &=& \frac{1}{v^2} \left(\Lambda^2 + m_A^2 - 2 m_C^2\right) \,, \label{e:lam4_2hdm}\\ 
		\lambda_5 &=&  \frac{1}{v^2} \left(\Lambda^2 - m_A^2\right) \,. \label{e:lam5_2hdm} 
	\end{eqnarray}
\label{e:lamb_mass_2hdm}
\end{subequations}

This set of equations allows us to trade the quartic parameters $\lambda_{i}$ for physical masses and mixing angles, thereby facilitating phenomenological analysis.

\subsection{Alignment limit in the 2HDM}
\label{sec:align_limit_2hdm}
The alignment limit in the 2HDM~\cite{Gunion:2002zf} is defined as the condition under which the CP-even neutral scalar denoted by $h$ in Eq.~\eqref{e:h125_2hdm} 
acquires tree-level couplings to SM gauge bosons and fermions that are identical to those of the SM Higgs boson. 
It can be easily shown that the combination~\cite{Das:2019yad,Bento:2017eti}
\begin{eqnarray}
		H_0 &=& \frac{1}{v}\left(v_1 h_1 + v_2 h_2\right) \equiv \cos \beta ~h_1 + \sin \beta ~h_2\,, 
		\label{e:H0state_def_2hdm}
\end{eqnarray}
can mimic the SM-Higgs couplings at the tree-level.
In view of this, the `Higgs basis' can be defined as~\cite{Georgi:1978ri,Lavoura:1994fv,Davidson:2005cw}
\begin{eqnarray}
	\begin{pmatrix}
		H_0 \\
		R 
	\end{pmatrix} &=& {\cal O}_\beta \begin{pmatrix}
	h_1 \\
	h_2 
\end{pmatrix}
\label{e:Higgs-basis}
\end{eqnarray}
These fields, in general, do not coincide with the physical eigenstates defined
in \Eqn{e:h125_2hdm}. Therefore, we may write
\begin{eqnarray}
		\begin{pmatrix}
		h \\
		H 
	\end{pmatrix} &=&  {\cal O} \begin{pmatrix}
		H_0 \\
		R 
	\end{pmatrix} \,,
	\label{e:h0_align_2hdm}
	\qquad {\rm with} \quad {\cal O} = {\cal O}_\alpha{\cal O}_\beta^T \,.
\end{eqnarray}
In the alignment limit, the physical eigenstate $h$  
completely overlaps with the $H_0$ state  which implies ${\cal O}_{11} = 1$.
In our parametrization,\footnote{In a more conventional notation~\cite{Gunion:2002zf}, the neutral scalar mixing angle defined in \Eqn{e:h125_2hdm}
	differs by $\pi/2$ and so the alignment limit is often expressed 
	as	 $\cos(\alpha - \beta) = 0$.}
this corresponds to~\cite{Das:2019yad}:
\begin{eqnarray}
	\label{e:alignment_2hdm}
\cos(\alpha - \beta) = 1 \,.
\end{eqnarray}
%
%
In passing, it is worth noting that the $2\times 2$ matrix proportional to 
$\frac{\Lambda^2}{v^2}$ in \Eqn{e:MS2} becomes diagonal in the Higgs basis defined in \Eqn{e:Higgs-basis}, as
\begin{eqnarray}
\frac{\Lambda^2}{v^2}\,{\cal O}_\beta 
\begin{pmatrix}
v_2^2  & -v_1 v_2 \\
- v_1 v_2 & v_1^2
\end{pmatrix}
{\cal O}_\beta^T 
= 
\begin{pmatrix}
0 & 0 \\
0 & \Lambda^2
\end{pmatrix}.
\end{eqnarray}
Consequently, in the alignment limit, the expression for $m_h^2$ is governed entirely by terms proportional to $v^2$, without any contribution from $\Lambda^2$. 
This observation is reassuring: if $h$ is to mimic the SM Higgs boson in the alignment limit, then -- consistent with the spirit of the SM -- its mass should arise purely from the electroweak vacuum expectation value, without contamination from the soft-breaking parameter. 
As we will demonstrate explicitly in the next section, the latter encapsulates the effects of nonstandard mass scales unrelated to the electroweak scale.

\section{Spontaneous generation of $m_{12}^2$ in a $Z_2$ invariant 2HDM extension}
\label{sec:Z2-2HDMS}

In this section, we demonstrate how the effects of the $m_{12}^2$ term can emerge from a UV-complete theory that possesses an exact $Z_2$ symmetry. 
The simplest way to illustrate this is by extending the 2HDM with a real $SU(2)_L$-singlet scalar field, $S$, which is odd under the $Z_2$ symmetry. 
The most general scalar potential consistent with these symmetry assignments can be written as,\footnote{It should be emphasized that the version of the 2HDM+singlet model considered here is structurally distinct from those commonly studied in the literature~\cite{Muhlleitner:2016mzt,Chen:2013jvg,Muhlleitner:2017dkd,Ferreira:2019iqb,Baum:2018zhf,Engeln:2020fld}. 
	In the latter models, the soft $Z_2$-breaking term is still present because the singlet field transforms trivially under this symmetry (although often non-trivially under an additional $Z_2$ symmetry) and thus cannot spontaneously break it. 
	In contrast, in our setup the $Z_2$ symmetry is respected by all terms in the scalar potential given in \Eqn{e:2HDMS}.}
\begin{eqnarray}
V(\Phi_{1}, \Phi_{2}, S) &=& 
m_{11}^{2}\,\Phi_{1}^{\dagger}\Phi_{1} 
+ m_{22}^{2}\,\Phi_{2}^{\dagger}\Phi_{2} 
+ \frac{\lambda_{1}}{2}\left(\Phi_{1}^{\dagger}\Phi_{1}\right)^{2} 
+ \frac{\lambda_{2}}{2}\left(\Phi_{2}^{\dagger}\Phi_{2}\right)^{2} \nonumber \\[4pt]
&& +\, \lambda_{3}\left(\Phi_{1}^{\dagger}\Phi_{1}\right)\left(\Phi_{2}^{\dagger}\Phi_{2}\right)
+ \lambda_{4}\left(\Phi_{1}^{\dagger}\Phi_{2}\right)\left(\Phi_{2}^{\dagger}\Phi_{1}\right)
+ \left[\frac{\lambda_{5}}{2}\left(\Phi_{1}^{\dagger}\Phi_{2}\right)^{2} + \text{h.c.}\right] \nonumber \\[4pt]
&& +\, m_{33}^{2}\,S^{2} 
+ \frac{\lambda_{s}}{8} S^{4} 
+ \frac{\lambda_{1s}}{2}\,\Phi_{1}^{\dagger}\Phi_{1}\,S^{2} 
+ \frac{\lambda_{2s}}{2}\,\Phi_{2}^{\dagger}\Phi_{2}\,S^{2}
- \left[\mu\,\Phi_{1}^{\dagger}\Phi_{2} + \text{h.c.}\right] S \,.
\label{e:2HDMS}
\end{eqnarray}
The new scalar field $S$ participates in the spontaneous breaking of the $Z_2$ symmetry. 
However, its VEV does not contribute to the electroweak  VEV, since $S$ is a singlet under the electroweak gauge group. 
Denoting the VEV of $S$ by $v_s$, the field can be expanded after SSB as
\begin{eqnarray}
S = v_s + h_s\,.
\end{eqnarray}
Since $v_s$ is unrelated to the electroweak VEV defined in \Eqn{e:2hdm_vev}, it can easily take values much larger than the electroweak scale.

Similar to the 2HDM case, the minimization conditions are used to trade the bilinears in terms of the VEVs as,
\begin{subequations}
	\begin{eqnarray}
		m_{11}^2 &=& \mu v_s \frac{v_2}{v_1} - \frac{1}{2}\left[\lambda_1 v_1^2 + \lambda_{1s} v_s^2 +\left
		(\lambda_3+\lambda_4+\lambda_5\right)v_2^2\right] \,, \\
		m_{22}^2 &=& \mu v_s \frac{v_1}{v_2} - \frac{1}{2}\left[\lambda_2 v_2^2  +  \lambda_{2s} v_s^2 + \left
		(\lambda_3+\lambda_4+\lambda_5\right)v_1^2\right]  \,. \\
		m_{33}^2 &=& \frac{1}{2} \mu v_1 v_2  - \frac{1}{4} \left[\lambda_{1s} v_1^2  +  \lambda_{2s} v_2^2 + \lambda_s v_s^2\right]  \,.
	\end{eqnarray}
	\label{e:minim_2hdms}
\end{subequations}

Next, we will discuss the mass matrices and 
the physical mass eigenstates in this extended model. To start with, the pseudo-scalar mass matrix is given by,
\begin{subequations}
	\begin{eqnarray}
		{\mathscr V}^{\rm mass}_P &=& \begin{pmatrix}
			z_1 & z_2 
		\end{pmatrix} \, \frac{{\mathscr M}_P^2}{2} \, \begin{pmatrix}
			z_1\\  z_2\\
		\end{pmatrix} \,,
	\label{e:PS_2hdms} \\
{\rm with,}\qquad		{\mathscr M}_P^2 &=& \begin{pmatrix}
			\mu v_s \frac{v_2}{v_1} - \lambda_5 v_2^2   & - \mu v_s + \lambda_5 v_1 v_2 \\
			- \mu v_s + \lambda_5 v_1 v_2 &	\mu v_s\frac{ v_1}{v_2} - \lambda_5 v_1^2 \\
		\end{pmatrix}  \\ 
		&=& \left[\frac{\mu v_s}{v_1 v_2} -\lambda_5 \right] \begin{pmatrix} v_2^2 & -v_1v_2 \\
			-v_1 v_2 & v_1^2 \label{e:mp_2hdms}\\
		\end{pmatrix} \,.
	\end{eqnarray}
	\label{e:mps_2hdms}
\end{subequations}
Similarly, the charged scalar mass matrix is given as,
\begin{subequations}
   \begin{eqnarray}
	{\mathscr M}_C^2 &=& \begin{pmatrix}
		\mu v_s \frac{v_2}{v_1} -\frac{1}{2} \left(\lambda_4+\lambda_5\right) v_2^2  &	-\mu v_s +\frac{1}{2} \left(\lambda_4+\lambda_5\right) v_1 v_2 \\
		-\mu v_s +\frac{1}{2} \left(\lambda_4+\lambda_5\right) v_1 v_2 &	\mu v_s\frac{ v_1}{v_2} -\frac{1}{2} \left(\lambda_4+\lambda_5\right)v_1^2 \\
	\end{pmatrix} \\ 
	&=& \left[\frac{\mu v_s}{v_1 v_2} -\frac{1}{2} \left(\lambda_4+\lambda_5\right)\right] \begin{pmatrix} v_2^2 & -v_1v_2 \\
		-v_1 v_2 & v_1^2 \label{e:mc_2hdms}\\
	\end{pmatrix} \,.
\end{eqnarray}
\label{e:mcs_2hdms}
\end{subequations}
Comparing \Eqs{e:mpsq}{e:mcsq} with \Eqs{e:mp_2hdms}{e:mc_2hdms}, we notice a striking similarity. Quite obviously, the mass
matrices in \Eqs{e:mp_2hdms}{e:mc_2hdms} can still be diagonalized as in \Eqn{e:mcmp_diag_2hdm} and the mass eigenvalues should be given by,
	\begin{eqnarray}\label{e:ma_mc_2hdms}
		m_{C}^2 =  - \frac{v^2}{2} (\lambda_4+\lambda_5) + \frac{\mu v_s}{\sin\beta \cos\beta} \, ,  \qquad \text{and} \qquad
		m_{A}^2 = - \lambda_5 v^2 + \frac{\mu v_s}{\sin\beta \cos\beta} \, .
	\end{eqnarray}
%
Comparison of \Eqn{e:ma_mc_2hdms} with its 2HDM counterpart in \Eqn{e:ma_mc_2hdm}
suggests that the 2HDM soft-breaking parameter should be directly related to the singlet VEV $v_s$ as,
\begin{eqnarray}
	m_{12}^2 = \mu v_s \equiv \Lambda^2 \sin\beta \cos\beta\,.
	\label{e:m12_mu_relation}
\end{eqnarray}

Unlike the pseudo-scalar and charged-scalar sectors, the CP-even scalar sector 
involves additional complexity. In particular, the neutral scalar sector now 
includes a singlet scalar field ($h_s$), which mixes with the doublet components. 
Accordingly, the $3\times3$ mass matrix in the CP-even sector can be written as
\begin{subequations}
	\begin{eqnarray}
	{\mathscr V}^{\rm mass}_S 
	= 
	\begin{pmatrix}
	h_1 & h_2 & h_s
	\end{pmatrix}
	\frac{{\mathscr M}_S^2}{2}
	\begin{pmatrix}
	h_1 \\[2pt] h_2 \\[2pt] h_s
	\end{pmatrix} \,,
	\end{eqnarray}
	where
	\begin{eqnarray}
	{\mathscr M}_S^2 =
	\begin{pmatrix}
	\lambda_1 v_1^2 +  \mu v_s \dfrac{v_2}{v_1} &
	(\lambda_3 + \lambda_4 + \lambda_5)v_1 v_2 - \mu v_s &
	\lambda_{1s} v_1 v_s - \mu v_2 \\[5pt]
	(\lambda_3 + \lambda_4 + \lambda_5)v_1 v_2 - \mu v_s &
	\lambda_2 v_2^2 + \mu v_s \dfrac{v_1}{v_2} &
	\lambda_{2s} v_2 v_s - \mu v_1 \\[5pt]
	\lambda_{1s} v_1 v_s - \mu v_2 &
	\lambda_{2s} v_2 v_s - \mu v_1 &
	\lambda_s v_s^2 + \mu v_s \dfrac{v_1 v_2}{v_s^2}
	\end{pmatrix} \,.
	\end{eqnarray}
	\label{e:cpeven_2hdms}
\end{subequations}
To rotate to the physical mass basis -- where the mass eigenstates are identified as 
$(h,\, H_1,\, H_2)$ with corresponding eigenvalues 
$(m_h,\, m_{H_1},\, m_{H_2})$ -- we introduce three mixing angles 
$\alpha$, $\epsilon$, and $\epsilon'$. 
The full $3\times3$ orthogonal rotation matrix ${\mathscr O}_\alpha$ 
is parametrized as a product of three sequential rotations:
\begin{subequations}
	\begin{eqnarray}
	{\mathscr O}_\alpha = \mathscr{R}_3 \cdot \mathscr{R}_2 \cdot \mathscr{R}_1 \,,
	\qquad \text{such that} \qquad
	\begin{pmatrix}
	h \\[2pt] H_1 \\[2pt] H_2
	\end{pmatrix}
	=
	{\mathscr O}_\alpha
	\begin{pmatrix}
	h_1 \\[2pt] h_2 \\[2pt] h_s
	\end{pmatrix} \,,
	\label{e:oalpha_2hdms}
	\end{eqnarray}
	where
	\begin{eqnarray} 
	\mathscr{R}_1 =
	\begin{pmatrix}
	\cos\alpha & \sin\alpha & 0 \\
	-\sin\alpha & \cos\alpha & 0 \\
	0 & 0 & 1
	\end{pmatrix}, 
	\quad
	\mathscr{R}_2 =
	\begin{pmatrix}
	\cos\epsilon & 0 & \sin\epsilon \\
	0 & 1 & 0 \\
	-\sin\epsilon & 0 & \cos\epsilon
	\end{pmatrix}, 
	\quad
	\mathscr{R}_3 =
	\begin{pmatrix}
	1 & 0 & 0 \\
	0 & \cos\epsilon' & \sin\epsilon' \\
	0 & -\sin\epsilon' & \cos\epsilon'
	\end{pmatrix}.
	\end{eqnarray}
	\label{e:rotcpeven_2hdms}
\end{subequations}
The diagonalization of the CP-even scalar mass matrix then reads:
\begin{eqnarray}\label{e:neutral_2hdms}
	\begin{pmatrix}
	m_h^2 & 0 & 0 \\[2pt]
	0 & m_{H_1}^2 & 0 \\[2pt]
	0 & 0 & m_{H_2}^2
	\end{pmatrix}
	=
	{\mathscr O}_\alpha \cdot {\mathscr M}_S^2 \cdot {\mathscr O}_\alpha^{T} \quad 
	\Rightarrow
	\quad
	{\mathscr O}_\alpha^{T} \cdot
	\begin{pmatrix}
	m_h^2 & 0 & 0 \\[2pt]
	0 & m_{H_1}^2 & 0 \\[2pt]
	0 & 0 & m_{H_2}^2
	\end{pmatrix}
	\cdot
	{\mathscr O}_\alpha
	=
	{\mathscr M}_S^2 \,.
\end{eqnarray}
Using \Eqs{e:ma_mc_2hdms}{e:m12_mu_relation} together with 
\Eqn{e:neutral_2hdms}, one can express all eight scalar quartic couplings 
($\lambda_{1,\dots,5}$, $\lambda_{1s}$, $\lambda_{2s}$, $\lambda_s$)
along with the bilinears $m_{11}^2$, $m_{22}^2$ and $m_{33}^2$ 
in terms of the physical parameters -- namely, the five scalar masses 
$(m_A,\, m_C,\, m_h,\, m_{H_1},\, m_{H_2})$, the three neutral mixing angles 
$(\alpha,\, \epsilon,\, \epsilon')$, $\Lambda$, and $\tan\beta$. 
We first display the explicit expressions for the five quartic parameters 
$\lambda_{1-5}$ appearing in the 2HDM potential of \Eqn{e:2hdm_pot} as follows:
\begin{subequations} 
	\begin{align}
		\lambda_1  & =  \frac{1}{v^2 \cos^2\beta} \left[m_h^2 \cos^2\alpha \cos^2 \epsilon + 
		m_{H_1}^2 \left(\cos \epsilon \sin\alpha + \cos \alpha \sin \epsilon \sin \epsilon^\prime\right)^2  \right. \nonumber \\
		 & \left.  +  m_{H_2}^2 \left(\cos \alpha \cos \epsilon^\prime \sin \epsilon - \sin \alpha \sin \epsilon^\prime \right)^2 \right] - \frac{\Lambda^2}{v^2}  \tan^2\beta \,,
		 \label{e:lam1_2hdms} \\ 
		\lambda_2 &= \frac{1}{v^2 \sin^2 \beta}  \left[m_h^2 \sin^2\alpha \cos^2 \epsilon + 
		m_{H_1}^2 \left(\cos \epsilon^\prime \cos \alpha - \sin \alpha \sin \epsilon \sin \epsilon^\prime\right)^2 \right. \nonumber \\ 
		& \left. + m_{H_2}^2 \left(\sin \alpha \cos \epsilon^\prime \sin \epsilon + \cos \alpha \sin \epsilon^\prime \right)^2 \right] - \frac{\Lambda^2}{v^2}  \frac{1}{\tan^2\beta} \,, \label{e:lam2_2hdms} \\ 
		\lambda_3 & =  \frac{1}{v^2}  \left[ 2 m_C^2 - \Lambda^2 + m_h^2 \cos^2 \epsilon \frac{\sin 2 \alpha}{\sin 2 \beta} -  m_{H_1}^2 \left(\frac{1}{2}\cos^2 \epsilon + \frac{3 }{4}\cos 2\epsilon^\prime - \frac{1}{4}\cos 2\epsilon \cos 2\epsilon^\prime \right) \frac{\sin 2 \alpha}{\sin 2 \beta}\right. \nonumber  \\ 
				& - \left. m_{H_2}^2 \left(\frac{1}{2}\cos^2 \epsilon - \frac{3 }{4}\cos 2\epsilon^\prime + \frac{1}{4}\cos 2\epsilon \cos 2\epsilon^\prime \right) \frac{ \sin 2 \alpha}{\sin 2\beta} +(m_{H_1}^2 - m_{H_2}^2) \sin \epsilon \sin 2\epsilon^\prime \frac{ \cos 2 \alpha}{\sin 2\beta}\right] \,,  \label{e:lam3_2hdms}\\ 
		\lambda_4 &= \frac{1}{v^2} \left(\Lambda^2 + m_A^2 - 2 m_C^2\right) \,, \label{e:lam4_2hdms} \\ 
		\lambda_5 &=  \frac{1}{v^2} \left(\Lambda^2 - m_A^2\right) \,. \label{e:lam5_2hdms}
	\end{align}
\label{e:lamb_mass_2hdms}
\end{subequations}
It is important to note that in the limit 
\begin{eqnarray}
	\epsilon \to 0 \,, \quad \epsilon^\prime \to 0 \,,
	\label{e:2hdm_limit}
\end{eqnarray} 
the above quartic couplings exactly 
match the relations given in \Eqn{e:lamb_mass_2hdm} for the 2HDM case. This correspondence arises because, when the mixing angles
$\epsilon$ and $\epsilon^\prime$ vanish, the singlet sector doesn't mix with the doublet components from the 2HDM spectrum.

For completeness, we also express the remaining quartic couplings $\lambda_{1s}\,, \lambda_{2s}\,,$ and  $\lambda_s$, in terms of the physical quantities:
\begin{subequations}
	\begin{eqnarray}
		\lambda_{1s} &=& \frac{\Lambda^2}{v_s^2}  \sin^2\beta + \frac{1}{2 v v_s \cos \beta} \left[(m_{H_2}^2 - m_{H_2}^2) \left(\sin\alpha \cos \epsilon \sin 2\epsilon^\prime - 
		\cos \alpha \cos^2 \epsilon^\prime  \sin 2 \epsilon \right)  \right.\nonumber \\ 
		&& + \left. (m_h^2 - m_{H_1}^2) \cos \alpha \sin 2 \epsilon \right] \,,  \\
		\lambda_{2s} &=& \frac{\Lambda^2}{v_s^2}  \cos^2\beta + \frac{1}{2 v v_s \sin \beta} \left[(m_{H_1}^2 - m_{H_2}^2) \left(\cos \alpha \cos \epsilon \sin 2\epsilon^\prime + 
		\sin \alpha \cos^2 \epsilon^\prime  \sin 2 \epsilon \right)  \right.\nonumber \\ 
		&& + \left. (m_h^2 - m_{H_1}^2) \sin \alpha \sin 2 \epsilon \right] \,, \\
	\lambda_s &=& \frac{v^2}{4 v_s^4} \Lambda^2 \sin^2 2\beta + \frac{1}{v_s^2}\left[m_h^2 \sin^2 \epsilon + \cos^2\epsilon \left(m_{H_2}^2 \cos^2\epsilon^\prime
	 + m_{H_1}^2 \sin^2\epsilon^\prime\right)\right] \,.
			\end{eqnarray}
		\label{e:singlet_quartics}
\end{subequations}
\subsection{Alignment limit in the $Z_2$-symmetric singlet-extended 2HDM}
Following Sec.~\ref{sec:align_limit_2hdm}, we can straightforwardly define the alignment limit in the full theory, namely in the $Z_2$-symmetric 2HDM extended by an additional scalar singlet. The definition of the SM-like state $H_0$ remains identical to that in \Eqn{e:H0state_def_2hdm}. However, owing to the presence of the singlet, the full rotation matrix ${\mathscr O}_\beta$ must now be a $3\times3$ matrix. The rotation that leads to $H_0$ is therefore given by
\begin{eqnarray}
\begin{pmatrix}
H_0 \\
R_1 \\
R_2
\end{pmatrix}
&=&
{\mathscr O}_{\beta}
\begin{pmatrix}
h_1 \\
h_2 \\
h_s
\end{pmatrix},
\qquad
{\mathscr O}_{\beta} =
\begin{pmatrix}
\cos\beta & \sin\beta & 0 \\
-\sin\beta & \cos\beta & 0 \\
0 & 0 & 1
\end{pmatrix}.
\label{e:obeta_3d}
\end{eqnarray}
The physical mass eigenstates are obtained through the rotation ${\mathscr O}_\alpha$ defined in \Eqn{e:rotcpeven_2hdms}. Combining \Eqs{e:rotcpeven_2hdms}{e:obeta_3d}, we can express the transformation between the physical and interaction bases as
\begin{eqnarray}
\begin{pmatrix}
h \\
H_1 \\
H_2
\end{pmatrix}
&=&
{\mathscr O}
\begin{pmatrix}
H_0 \\
R_1 \\
R_2
\end{pmatrix},
\qquad
{\mathscr O} = {\mathscr O}_\alpha\, {\mathscr O}_{\beta}^T.
\label{e:h0_align_2hdms}
\end{eqnarray}
Once again, to ensure that $H_0$ coincides entirely with the SM-like Higgs $h$, we require ${\mathscr O}_{11} = 1$, which implies
\begin{eqnarray}
\cos(\alpha - \beta)\, \cos\epsilon = 1 \,.
\label{e:O11_2hdms}
\end{eqnarray}
Hence, the {\it alignment condition} in the full theory can be expressed as
\begin{eqnarray}
\cos(\alpha - \beta) = 1,
\qquad
\epsilon = 0.
\label{e:alignment1_2hdms}
\end{eqnarray}
It is worth noting that even in this {alignment limit}, the states $R_1$ and $R_2$ can still mix via the angle $\epsilon'$, giving rise to the physical eigenstates $H_1$ and $H_2$. Consequently, due to this residual mixing, neither $H_1$ nor $H_2$ can be directly identified with the heavy Higgs eigenstate $H$ originating from the pure 2HDM spectrum.  
To recover the {pure 2HDM alignment limit}, where $H_1$ can be identified with $H$, one must additionally impose
\begin{eqnarray}
\epsilon' = 0 \,.
\label{e:alignment2_2hdms}
\end{eqnarray}
In view of this distinction, it may be reasonable to designate \Eqn{e:alignment1_2hdms} as the {\em Higgs alignment limit}, and \Eqn{e:2hdm_limit} as the {\em 2HDM limit}.

\subsection{From soft-breaking to singlet VEV: understanding the origin of non-standard scalar masses} 

We have previously shown that comparing the pseudoscalar and charged scalar masses in the minimal 2HDM and its singlet-extended counterpart leads to a relation
between the
soft-breaking parameter $m_{12}^2$, the trilinear coupling $\mu$ and the singlet VEV $v_s$, or
equivalently to $\Lambda^2$, as
given in \Eqn{e:m12_mu_relation}. It is important to emphasize that the non-electroweak contributions to the pseudoscalar and charged scalar masses in \Eqn{e:ma_mc_2hdm} arise from the term proportional to the soft-breaking parameter $m_{12}^2$. In the singlet-extended scenario,
as shown in \Eqn{e:ma_mc_2hdms} this 
notion of non-electroweak source of mass is accentuated, as $\Lambda^2$ becomes directly related to the singlet VEV $v_s$, which doesn't contribute to the EWSB.

Thus, the origin of the additional mass for the non-standard CP-odd and charged scalars is directly tied to the high-scale VEV of the singlet scalar in the extended theory. 
In other words, the UV completion of the softly-broken $Z_2$-symmetric 2HDM can be understood as emerging from a 
perfectly $Z_2$-symmetric 2HDM extended by a singlet scalar whose VEV {combined with the other nonstandard parameters can mimic the effect of $m_{12}^2$.}

To understand the correspondence in the neutral scalar sector, it is instructive to compare the inverted relations between the quartic couplings and physical parameters in the minimal 2HDM and its singlet-extended version. According to \Eqn{e:lamb_mass_2hdm}, the quartic couplings relevant to the CP-even sector in the minimal 2HDM are 
$\lambda_1\,,\lambda_2$ and $\lambda_3$. 
Now, by imposing the conditions from Eqs.(\ref{e:alignment1_2hdms}) and (\ref{e:alignment2_2hdms}) into the extended scenario's expressions -- Eqs.~(\ref{e:lam1_2hdms}), (\ref{e:lam2_2hdms}), and (\ref{e:lam3_2hdms})-- one finds that the resulting quartic couplings exactly match {the corresponding expressions in Eqs.~(\ref{e:lam1_2hdm}), (\ref{e:lam2_2hdm}), and (\ref{e:lam3_2hdm})}.
This confirms our expectation that,
to decouple the nonstandard scalars from the electroweak scale by making them heavy, we must require $v_s \gg v$. This is to say that, if the nonstandard  masses are to be super heavy -- well beyond the TeV 
	scale -- they better derive their masses almost entirely from the non-electroweak sources (in this case, $v_s$).


It is important to note that, although an $SU(2)_L$-singlet scalar does not contribute to the electroweak VEV, it can affect the mass relations between the $W$- and $Z$-bosons at the loop level~\cite{Grimus:2007if, Grimus:2008nb} through the mixing with the components
of the doublets.  
	As such, we find it useful to add some remarks on the $T$-parameter for this case.  
	A more general treatment can be found in Appendix~\ref{sec:custodial}, whereas here we keep to the relevant case at hand.

	From Appendix~\ref{sec:custodial}, we see that the conditions for $\Delta T_\text{NP}=0$ are the alignment limit, having no mixing between the pseudoscalars arising from $\Phi_1$ and $\Phi_2$, and those arising from singlets, as well as a perfect match between the charged- and pseudo-scalar mass matrices.  
	However, the singlet in this case is real, and so, given the assumption of CP-conservation, the second condition is trivially satisfied, as there exist no other pseudoscalars.  
	From \Eqs{e:mp_2hdms}{e:mc_2hdms}, we see that the mixing is necessarily the same (as there is only one physical eigenstate), but the degeneracy of the masses requires $\lambda_4=\lambda_5$, which we can cross-check with \Eqn{e:lamb_mass_2hdms} that is achievable through $m_A=m_C$.  
	Finally, the alignment limit can be read from \Eqn{e:alignment1_2hdms}.  
	As such, we see that none of these conditions place any restriction of $\epsilon'$, which agrees with the results found in~\cite{Li:2025zga}.  
	This result is also to be expected when we think about the arrangement of the fields into custodial multiplets~\cite{Kundu:2021pcg, Das:2022gbm}.  
	More specifically, since the Goldstones are in a custodial triplet, we must have the remaining pseudoscalars and charged-fields coming from the doublets in custodial triplets, which (here) implies $\mathcal{M}_P^2=\mathcal{M}_C^2$.  
	On the other hand, the CP-even fields are custodial singlets, and so these should be allowed to mix without a significant impact on the theory.  
	The only condition here comes in the form of the alignment limit, such that the contributions from Eq.~\eqref{eq:part5} cancel exactly that coming from Eq.~\eqref{eq:part6}.
	

\subsection{Phenomenological insights for physics beyond the 2HDM}


The analysis above prompts an intriguing question: Suppose that, in {a wishful} future, 
the particle spectrum predicted by the 2HDM is observed {in the experiments}, and all 
couplings involving only the 2HDM particles have been precisely measured. 
In such a scenario, how can we identify the existence of physics beyond the 2HDM?
More specifically, how can we detect the presence of nonzero values of $\epsilon^\prime$, as defined in Eq.~(\ref{e:rotcpeven_2hdms}).

To explore this, we will start by examining the constraints imposed by unitarity and their implications for physics beyond the 2HDM. 
Consider first the scattering process $W^+ W^-  \to W^+ W^-$. In the SM, the diagram mediated by the neutral 
scalar Higgs boson is essential for canceling the energy growth arising from the quartic gauge coupling $g_{WWWW}$.
When the scalar sector is extended beyond the SM, the interaction Lagrangian describing the trilinear couplings between 
the neutral scalars and the gauge bosons can be written as,
%
\begin{eqnarray}
	{\cal L}_{\rm int}^{W^+W^-S^0} = g M_W \left( \kappa_W^h h + \kappa_W^{H_1} H_1 + \kappa_W^{H_2} H_2 \right) W_\mu^+ W^{\mu -} \,,
\end{eqnarray}
where, $\kappa_W^S,  \{S = h \,, H_1\,, H_2\}$ denotes the gauge couplings of the respective scalar $S$
to $W$ normalized to the SM Higgs coupling: 
\begin{subequations}
\begin{eqnarray}
	\kappa_W^h &=& \cos(\alpha - \beta) \cos \epsilon \,, \\
	\kappa_W^{H_1} &=& -\sin(\alpha - \beta) \cos \epsilon^\prime - \cos(\alpha -\beta)\sin\epsilon \sin\epsilon^\prime \,, \\
	\kappa_W^{H_2} &=& \sin(\alpha - \beta) \sin \epsilon^\prime - \cos(\alpha -\beta)\sin\epsilon \cos \epsilon^\prime \,.
\end{eqnarray}
\label{e:kappa_W_S}
\end{subequations}
Therefore, to ensure the cancellation of the energy growth from the quartic gauge interaction, the following sum rule must hold:\cite{Gunion:1990kf}
\begin{eqnarray}
\left(\kappa_W^h\right)^2  + \left(\kappa_W^{H_1}\right)^2 + \left(\kappa_W^{H_2}\right)^2 & = & 1 \,. 
\label{e:sum_rule_1}
\end{eqnarray}
It can be easily checked that the gauge coupling modifiers given in Eq.~(\ref{e:kappa_W_S}) indeed satisfy the above sum rule.
Now, assuming only $h$ and $H_1$ have been found at the experiment, the existence of $H_2$ can be ascertained by 
checking whether following combination vanishes:
\begin{eqnarray}
	1 - \left(\kappa_W^h\right)^2 -  \left(\kappa_W^{H_1}\right)^2 = \left(\sin(\alpha - \beta) \sin \epsilon^\prime - \cos(\alpha -\beta)\sin\epsilon \cos \epsilon^\prime\right)^2 \,.
\end{eqnarray} 
It is evident that this combinantion vanishes when there is no mixing between the 2HDM sector and 
the singlet sector. But, more importantly for nonzero mixing,
this sum rule can only reveal the presence of a nonzero $\epsilon^\prime$ if the coupling of the
scalar $h$ deviates from that of the SM Higgs. In the alignment limit, given in Eq.~(\ref{e:alignment1_2hdms}),
i.e. $\cos(\alpha - \beta) = 1 \,, \epsilon = 0$, the sum rule is automatically satisfied regardless of 
the value of $\epsilon^\prime$. 

Therefore, a more intriguing question arises: Can we detect signs of physics beyond the
2HDM even if nature favors the alignment limit? This requires some creative insights but following the same reasoning as before, the answer remains positive. 
The key, in this case, is to involve certain 2HDM particles as external states in the scattering processes.

For example, we consider the scattering $W_L^+ W_L^- \to H^+ H^-$. The relevant interaction Lagrangian describing this
process is given as,
\begin{subequations}
\begin{eqnarray}
	{\cal L}_{\rm int}^{W^\pm S^\mp S^0} &=& \dfrac{i g}{2} \left[ \kappa_{W^+ H^-}^h \left\{\left(\partial^\mu H^-)h - H^-(\partial^\mu h\right) \right\} W^+_\mu + \kappa_{W^+ H^-}^{H_1} \left\{\left(\partial^\mu H^-)H_1 - H^-(\partial^\mu H_1\right) \right\} W^+_\mu \right. \, \nonumber \\
	&& \left. + \kappa_{W^+ H^-}^{H_2} \left\{\left(\partial^\mu H^+)H_2 - H^+(\partial^\mu H_2\right) \right\} W^+_\mu + h.c.\right] \, , \\
	{\cal L}_{\rm int}^{W W S^+ S^-} &=& \dfrac{g^2}{2} W^+_\mu W^-_\mu H^+ H^-  \,.
\label{e:lag_w_hpm_s0}
\end{eqnarray}
\end{subequations}
where, the couplings $\kappa^{S^0}_{W^+ H^-}\,, (S_0 \equiv h\,, H_1\,, H_2)$ are given as follows:
\begin{subequations}
	\begin{eqnarray}
		\kappa_{W^+ H^-}^h &=& \cos \alpha \sin \beta \cos \epsilon  + \cos \beta \cos \alpha \sin \epsilon \sin \epsilon^\prime + \cos \beta \sin \alpha \cos \epsilon^\prime \,, \\
	    \kappa_{W^+ H^-}^{H_1} &=& - \cos \alpha \cos \beta \cos \epsilon^\prime + \sin \alpha \sin \beta \cos \epsilon + \sin \alpha \cos \beta \sin \epsilon \sin \epsilon^\prime  \,, \\
    	\kappa_{W^+ H^-}^{H_2} &=& \sin \beta \sin \epsilon - \cos \beta \cos \epsilon \sin \epsilon^\prime \,.
	\end{eqnarray}
\label{e:kappa_w_hpm_s0}
\end{subequations}
Here, the relevant sum rule that must hold to cancel the energy growth is given as\footnote{The s-channel neutral higgs ($h$) mediated diagram will not have energy growths in their
amplitudes. Also, note that the $W^+_\mu Z^\mu H^-$ interaction does not exist in this model.}\cite{Gunion:1990kf},
\begin{eqnarray}
	\left(\kappa_{W^+ H^-}^h\right)^2 + \left(\kappa_{W^+ H^-}^{H_1}\right)^2 + \left(\kappa_{W^+ H^-}^{H_2}\right)^2 = 1\,.
	\label{e:sum_rule_2}
\end{eqnarray}
Again, it is straightforward to verify that the couplings modifiers given in Eq.~(\ref{e:kappa_w_hpm_s0}) satisfy 
the above sum rule outlined above. Therefore, following a similar approach as before, in the anticipated scenario where 
the couplings of both $h$ and $H_1$ 
are known, the existence of $H_2$ can be inferred by examining the following expression:
\begin{eqnarray}
	1 - \left(\kappa_{W^+ H^-}^h\right)^2 - \left(\kappa_{W^+ H^-}^{H_1}\right)^2 = \left(\sin \beta \sin \epsilon - \cos \beta \cos \epsilon \sin \epsilon^\prime\right)^2 \,.
	\label{e:sum_rule_h2}
\end{eqnarray}

It is noteworthy that even in the alignment limit defined by \Eqn{e:alignment1_2hdms}, 
the right-hand side of \Eqn{e:sum_rule_h2} does not vanish; instead, it equals $\left(\cos \beta \sin \epsilon^\prime \right)^2$. 
Consequently, any nonzero deviation in the expression on the left of \Eqn{e:sum_rule_h2} would 
imply a nonzero $\epsilon^\prime$, thereby indicating the presence of physics
beyond the 2HDM.
\section{2HDMs with a softly-broken continuous symmetry}
\label{sec:U1-2HDM}

For 2HDMs with a softly-broken continuous symmetry, we need to be particularly careful while constructing the UV complete
version. This is because the continuous symmetry can not be extended directly to the full theory as the spontaneous
breaking of the symmetry will lead to a massless scalar if we do so.  One easy way to circumvent this is to ensure that softly-broken continuous symmetries at the electroweak scale
arise as accidental remnants of a spontaneously broken discrete symmetry at a high scale. 

To give an explicit example, let us consider the 2HDM with softly-broken $U(1)$ symmetry, the scalar potential of which is simply given by \Eqn{e:2hdm_pot}
with $\lambda_5 = 0$. The expression of \Eqn{e:2hdm_pot}, in this case, will reduce to,
\begin{subequations}
	\begin{eqnarray}
		\lambda_1 &=& 	\frac{1}{v^2 \cos^2\beta} \left(m_h^2 \cos^2\alpha + 
		m_H^2 \sin^2 \alpha\right) - \frac{\Lambda^2}{v^2}  \tan^2\beta \,; \label{e:lam1_u1_2hdm}\\
		\lambda_2 &=& \frac{1}{v^2 \sin^2 \beta} \left(m_H^2 \cos^2 \alpha +  
		m_h^2 \sin^2 \alpha\right) - \frac{\Lambda^2}{v^2}  \frac{1}{\tan^2 \beta} \,; \label{e:lam2_u1_2hdm}\\
		\lambda_3 &=& \frac{1}{v^2} \left(2 m_C^2 - \Lambda^2 - \left(m_H^2 - m_h^2\right) \frac{
			\cos \alpha  \sin \alpha}{\cos \beta \sin \beta}\right)\,; \label{e:lam3_u1_2hdm} \\ 
		\lambda_4 &=& \frac{2}{v^2} \left(\Lambda^2 -  m_C^2\right) \,; \label{e:lam4_u1_2hdm} \\ 
		\lambda_5 &=&  0 \,; \label{e:lam5_u1_2hdm} 
	\end{eqnarray}
	\label{e:lamb_mass_u12hdm}
\end{subequations}
where, we should keep in mind that
\begin{eqnarray}
	 \Lambda^2 \equiv m_A^2 \,.
	 \label{e:u1mA}
\end{eqnarray}
\subsection{A UV complete framework: $Z_3$-symmetric 2HDM with a complex singlet}
For a simple UV-complete scenario that leads to the 2HDM with softly-broken $U(1)$
symmetry, we extend the model with an additional complex scalar singlet (under $SU(2)_L \times U(1)_Y$) $\mathbb{S}$. 
Additionally, we impose a {\bf $Z_3$}-symmetry under which the scalar fields transform as follows:\cite{Heinemeyer:2021msz}
\begin{eqnarray}
	\Phi_1 \to \Phi_1 \,, \quad \Phi_2 \to \omega \Phi_2 \,, \quad {\mathbb S} \to \omega^2 {\mathbb S}
\end{eqnarray}
Therefore, the most general scalar potential obeying these symmetries will be given by, 
\begin{eqnarray}
	V_{Z_3}(\Phi_{1}, \Phi_{2},{\mathbb S}) & = & m^{2}_{11}\Phi^{\dagger}_{1}\Phi_{1} +  m^{2}_{22}\Phi^{\dagger}_{2}\Phi_{2} + \frac{\lambda_{1}}{2}\left(\Phi^{\dagger}_{1}\Phi_{1}\right)^{2} + \frac{\lambda_{2}}{2}\left(\Phi^{\dagger}_{2}\Phi_{2}\right)^{2}  \nonumber \\ 
	&& +\, \lambda_{3}\left(\Phi^{\dagger}_{1}\Phi_{1}\right)\left(\Phi^{\dagger}_{2}\Phi_{2}\right) + \lambda_{4}\left(\Phi^{\dagger}_{1}\Phi_{2}\right)\left(\Phi^{\dagger}_{2}\Phi_{1}\right) 
	+  m^{2}_{33}{\mathbb S}^*{\mathbb S} + \frac{\lambda_s}{8} \left({\mathbb S}^*{\mathbb S}\right)^2 
	 \nonumber  \\ 
	&& +\,\frac{\lambda_{1s}}{2} \left(\Phi^{\dagger}_{1}\Phi_{1}\right) \left({\mathbb S}^*{\mathbb S}\right) +  \frac{\lambda_{2s}}{2} \left(\Phi^{\dagger}_{2}\Phi_{2}\right) \left({\mathbb S}^*{\mathbb S}\right)
	- \left(\mu \Phi^{\dagger}_{1}\Phi_{2} {\mathbb S} +\text{h.c.} \right)
	\nonumber \\
	&& -\,\left(\lambda_{4s} \Phi^{\dagger}_{2}\Phi_{1} {\mathbb S} {\mathbb S} +\text{h.c.}  \right) 
	- \left( \mu_s {\mathbb S}^3 +\text{h.c.}  \right) \,.
	\label{e:u1_2hdms_pot}     
\end{eqnarray}%
The additional complex scalar field can develop a nonzero VEV, and be expanded as:
\begin{eqnarray}
	{\mathbb S} = v_s + \frac{1}{\sqrt{2}}\left(h_s + i z_s\right) \,.
\end{eqnarray}
The minimization conditions can therefore be written as,
\begin{subequations}
	\begin{eqnarray}
		m_{11}^2 &=& \left(\mu v_s + \lambda_{4s} v_s^2 \right) \frac{v_2}{v_1} - \frac{1}{2}\left[\lambda_1 v_1^2 + \lambda_{1s} v_s^2 +\left
		(\lambda_3+\lambda_4\right)v_2^2\right] \,, \\
		m_{22}^2 &=& \left(\mu v_s + \lambda_{4s} v_s^2 \right) \frac{v_1}{v_2} - \frac{1}{2}\left[\lambda_2 v_2^2  +  \lambda_{2s} v_s^2 + \left
		(\lambda_3+\lambda_4\right)v_1^2\right]  \,, \\
		m_{33}^2 &=& \frac{1}{2} \left(\mu + 2 \lambda_{4s} v_s \right) \frac{v_1 v_2}{v_s}  - \frac{1}{4} \left[\lambda_{1s} v_1^2  +  \lambda_{2s} v_2^2 + \lambda_s v_s^2 - 12 \mu_s v_s \right]  \,.
	\end{eqnarray}
	\label{e:minim_2hdms_u1}
\end{subequations}
 The charged scalar mass matrix can be obtained as,
\begin{subequations}
	\begin{eqnarray}
		{\mathbb M}_C^2 &=& \begin{pmatrix}
			\left(\mu v_s + \lambda_{4s}v_s^2\right) \frac{v_2}{v_1} -\frac{1}{2} \lambda_4 v_2^2  &	-\left(\mu v_s + \lambda_{4s}v_s^2\right) + \frac{1}{2} \lambda_4 v_1 v_2 \\
			-\left(\mu v_s + \lambda_{4s}v_s^2\right) + \frac{1}{2} \lambda_4 v_1 v_2 &	\left(\mu v_s + \lambda_{4s}v_s^2\right)\frac{ v_1}{v_2} -\frac{1}{2} \lambda_4v_1^2 \\
		\end{pmatrix} \\ 
		&=& \left[\frac{\left(\mu v_s + \lambda_{4s}v_s^2\right)}{v_1 v_2} -\frac{1}{2} \lambda_4\right] \begin{pmatrix} v_2^2 & -v_1v_2 \\
			-v_1 v_2 & v_1^2 \label{e:mc_2hdms_u1}\\
		\end{pmatrix} \,.
	\end{eqnarray}
	\label{e:mcs_2hdms_u1}
\end{subequations}
Once again, this matrix is
diagonalized by the rotation ${\cal O}_\beta$ given in Eq.~(\ref{e:Obeta}), with the charged-Higgs mass given by
\begin{eqnarray}
	m_{C}^2 &=&  \frac{\mu v_s}{\sin\beta \cos\beta} + \frac{ \lambda_{4s}v_s^2}{\sin\beta \cos\beta} -  \lambda_4 \frac{v^2}{2} \,.
\label{e:mc2_2hdms_u1}
\end{eqnarray}
We can compare this expression with the corresponding one in Eq.~(\ref{e:mc_2hdm}), in the limit $\lambda_5 \to 0$,
and identify the soft-symmetry breaking parameter $m_{12}^2$ in the following way:
\begin{subequations}
\begin{eqnarray}
	m_{12}^2 = \left(\mu v_s + \lambda_{4s}v_s^2\right) \equiv \left(\Lambda_1^2 + \Lambda_2^2\right) \sin \beta \cos \beta \,,
	\label{e:softbreaking_2hdms_u1} 
\end{eqnarray}
where, we have introduced
\begin{eqnarray}
	\Lambda_1^2 \equiv \frac{\mu v_s}{\sin \beta \cos \beta} \,, \quad \Lambda_2^2 \equiv \frac{\lambda_{4s} v_s^2}{\sin\beta \cos\beta} \,.
\label{e:soft_Lam1_Lam2_u1}
\end{eqnarray}
\end{subequations}
The pseudoscalar mass matrix differs from previous cases due to the presence of a complex scalar singlet, which allows the CP-odd component to mix with the CP-odd partners of the doublets.
The pseudoscalar mass matrix is now a $3\times 3$ matrix and is given by,
\begin{eqnarray}
{\mathbb M}_p^2 &=& \begin{pmatrix}
		\left(\mu v_s + \lambda_{4s}v_s^2\right) \frac{v_2}{v_1} & -\left(\mu v_s + \lambda_{4s}v_s^2\right) & 
		\frac{v_2}{\sqrt{2}} \left( - \mu + 2\lambda_{4s}v_s\right) \\ 
		- \left(\mu v_s + \lambda_{4s}v_s^2\right) & \left(\mu v_s + \lambda_{4s}v_s^2\right) \frac{v_1}{v_2} &
		\frac{v_1}{\sqrt{2}} \left(\mu - 2 \lambda_{4s}v_s\right) \\
		\frac{v_2}{\sqrt{2}} \left( - \mu + 2\lambda_{4s}v_s\right) & \frac{v_1}{\sqrt{2}} \left(\mu - 2 \lambda_{4s}v_s\right) & 9 v_s \mu_s + \frac{v_1 v_2 }{4 v_s}\left(\mu + 4 \lambda_{4s}v_s\right)
	\end{pmatrix}.
\label{e:mps_2hdms_u1}
\end{eqnarray}
This matrix can be block diagonalized by the $3\times 3$  matrix ${\mathscr O}_{\beta}$ defined in Eq.~(\ref{e:obeta_3d}) as follows:
\begin{eqnarray}
	{\mathbb B}_P^2 \equiv {\mathscr O}_{\beta} \cdot {\mathbb M}_P^2 \cdot {\mathscr O}_{\beta}^T = \begin{pmatrix}
		0 & 0 & 0 \\
		0 & \frac{\mu v_s + \lambda_{4s}v_s^2}{\sin \beta \cos \beta} & \frac{v}{\sqrt{2}}\left(\mu - 2 v_s \lambda_{4s}\right) \\
		0 & \frac{v}{\sqrt{2}}\left(\mu - 2 v_s \lambda_{4s}\right) & 9 \mu_s v_s + \frac{v^2 \sin 2\beta}{4 v_s} \left(\mu + 4 v_s \lambda_{4s}\right) \\
	\end{pmatrix}.
\label{e:mp_bp_2hdms_u1}
\end{eqnarray}
Following this, the pseudoscalar mass matrix can be fully diagonalized by an additional orthogonal rotation ${\cal O}_\delta$ defined as
\begin{subequations}
	\begin{eqnarray}
	{\cal O}_\delta = \begin{pmatrix}
	1 & 0 & 0 \\
	0 & \cos \delta & \sin \delta \\
	0 & - \sin \delta & \cos \delta \\
	\end{pmatrix}.
	\end{eqnarray}
	Therefore, the physical mass eigenstates can be expressed as
	\begin{eqnarray}
	\begin{pmatrix}
	G_0 \\
	A_1 \\
	A_2 \\
	\end{pmatrix} = {\cal O}_\delta {\mathscr O}_\beta 	\begin{pmatrix}
	z_1 \\
	z_2 \\
	z_s \\
	\end{pmatrix} .
	\end{eqnarray}
\label{e:Odelta}
\end{subequations}
The ${\cal O}_\delta$ transformation diagonalizes the nonzero $2\times 2$ submatrix in ${\mathbb B}_P^2$ into a diagonal matrix where the
diagonal entries refer to the mass eigenvalues of the two physical pseudoscalars $A_1$
and $A_2$. Therefore, one may write:
\begin{eqnarray}
{\cal O}_\delta \cdot {\mathbb B}_P^2 \cdot {\cal O}_\delta^T = {\rm diag} \left(0\,, m_{A_1}^2\,, m_{A_2}^2\right) .
\end{eqnarray}	
Inverting the above relation, the corresponding entries of ${\mathbb B}_P^2$ can be expressed as,
\begin{subequations}
\begin{eqnarray}
	\left({\mathbb B}_P^2 \right)_{22} & \equiv & m_{A_1}^2 \cos^2 \delta + m_{A_2}^2 \sin^2 \delta = \Lambda_1^2 + \Lambda_2^2 \,, \label{e:L1L2} \\
	\left({\mathbb B}_P^2 \right)_{23} & \equiv & \left(m_{A_1}^2 - m_{A_2}^2 \right) \cos \delta \sin \delta = \frac{v}{\sqrt{2} v_s} \left(\Lambda_1^2 - 2 \Lambda_2^2\right) \sin\beta \cos\beta \,, \\
	\left({\mathbb B}_P^2 \right)_{33} & \equiv & m_{A_1}^2 \sin^2 \delta + m_{A_2}^2 \cos^2 \delta = 9 \mu_s v_s + \frac{v^2}{2 v_s^2} \sin^2 \beta \cos^2 \beta \left( \Lambda_1^2 + 4 \Lambda_2^2 \right) \,,
\end{eqnarray}
\label{e:bps2_2hdms_u1} 
\end{subequations}
where $\mu$ and $\lambda_{4s}$ have been substituted using the definitions in 
\Eqn{e:soft_Lam1_Lam2_u1}. Using the relations in \Eqn{e:bps2_2hdms_u1},
the potential parameters  $\mu$, $\mu_s$ and $\lambda_{4s}$ or equivalently,
$\Lambda_1^2$, $\Lambda_2^2$ and $\mu_s$ can be expressed in terms of $m_{A_1}^2$,
$m_{A_2}^2$ and $\delta$. Among the relations in \Eqn{e:bps2_2hdms_u1}, 
\Eqn{e:L1L2} is particularly interesting because in the limit $\delta\to 0$, it reduces to
\begin{eqnarray}
\Lambda_1^2 + \Lambda_2^2 = m_{A_1}^2 \,.
\label{e:LL1L2}
\end{eqnarray}
From \Eqn{e:Odelta} we can easily see that the pseudoscalar $A_1$ is defined
purely in terms of the doublet fields in the limit $\delta\to 0$. Thus, it is
quite natural to associate $A_1$ with the pseudoscalar ($A$) from 2HDM in the limit
$\delta\to 0$. Therefore, \Eqn{e:LL1L2} should be compared with \Eqn{e:u1mA} and
consequently, we should conclude
\begin{eqnarray}
	\Lambda^2 = \Lambda_1^2 + \Lambda_2^2 \,.
	\label{e:L1+L2}
\end{eqnarray}
With such an identification, \Eqn{e:mc2_2hdms_u1} becomes consistent with \Eqn{e:lam4_u1_2hdm}.

The analysis of the CP-even sector closely parallels that of the $Z_2$- symmetric 2HDM extended by a singlet. The mass squared matrix in the CP-even sector is again a $3\times 3$ matrix as follows:
\begin{subequations}
\begin{eqnarray}
	{\mathbb M}_S^2 =  \begin{pmatrix}
	\left({\mathbb M}_S^2\right)_{11}	  &	\left({\mathbb M}_S^2\right)_{12} 	&  \left({\mathbb M}_S^2\right)_{13} \\
	\left({\mathbb M}_S^2\right)_{12}   & 	\left({\mathbb M}_S^2\right)_{22}	&	\left({\mathbb M}_S^2\right)_{23}  \\
		\left({\mathbb M}_S^2\right)_{13}	 &	 \left({\mathbb M}_S^2\right)_{23}	&  \left({\mathbb M}_S^2\right)_{33} \\
	\end{pmatrix},
\label{e:cpeven_2hdms_u1}
\end{eqnarray}
where,
\begin{eqnarray}
\left({\mathbb M}_S^2\right)_{11} &=& \lambda_1 v_1^2 +  \left(\mu v_s + \lambda_{4s}v_s^2\right) \frac{v_2}{v_1} \,,\\
\left({\mathbb M}_S^2\right)_{12} &=& (\lambda_3+\lambda_4) v_1 v_2 -\left(\mu v_s + \lambda_{4s}v_s^2\right) \,, \\
\left({\mathbb M}_S^2\right)_{13} &=& \frac{1}{\sqrt{2}}\left\{\lambda_{1s} v_s v_1 - ( 2 \lambda_{4s} v_s + \mu ) v_2 \right\} \,, \\
\left({\mathbb M}_S^2\right)_{22} &=&  \lambda_2 v_2^2  + \left(\mu v_s + \lambda_{4s}v_s^2\right) \frac{v_1}{v_2}  \,, \\
\left({\mathbb M}_S^2\right)_{23} &=& \frac{1}{\sqrt{2}}\big\{ \lambda_{2s} v_s v_1 - (2 \lambda_{4s}v_s + \mu) v_2\big\} \,, \\
\left({\mathbb M}_S^2\right)_{33} &=& \frac{1}{2}\lambda_s v_s^2 +  \frac{\mu}{2 v_s}v_1 v_2 - 3 \mu_s v_s \,.
\end{eqnarray}
\end{subequations}
We identify the mass eigenstates as $(h\,, H_1\,, H_2)$ with corresponding masses $(m_h\,, m_{H_1} \,, m_{H_2})$ in this scenario. These states are obtained via the rotation $\mathscr{O}_\alpha$ as defined in Eq.~(\ref{e:oalpha_2hdms}).

Following a similar procedure, the quartic couplings can be expressed in terms of the physical masses and mixing parameters. These include the six scalar masses ($ m_C\,, m_{A_1}\,, m_{A_2}\,, m_h\,, m_{H_1}\,, m_{H_1}$), the four neutral mixing angles ($\alpha\,, \epsilon\,,\epsilon^\prime, \delta$) and $\tan\beta$. However, using the relation provided in Eq.~(\ref{e:soft_Lam1_Lam2_u1}), the pseudoscalar masses can be traded for the soft-breaking parameters $\Lambda_1$ and $\Lambda_2$.
\begin{subequations} 
	\begin{eqnarray}
		\lambda_1 &=& 	\frac{1}{v^2 \cos^2\beta} \left[m_h^2 \cos^2\alpha \cos^2 \epsilon + 
		m_{H_1}^2 \left(\cos \epsilon^\prime \sin \alpha - \cos \alpha \sin \epsilon \sin \epsilon^\prime \right)^2 \right. \nonumber \\ 
		&& \left. + m_{H_2}^2 \left(\cos \alpha \cos \epsilon^\prime \sin \epsilon - \sin \alpha \sin \epsilon^\prime \right)^2 \right] - \frac{\Lambda^2}{v^2}  \tan^2\beta \,,
		\label{e:lam1_2hdms_u1} \\ 
		\lambda_2 &=& \frac{1}{v^2 \sin^2 \beta} \left[m_h^2 \sin^2\alpha \cos^2 \epsilon + 
		m_{H_1}^2 \left(\cos \epsilon^\prime \cos \alpha - \sin \alpha \sin \epsilon \sin \epsilon^\prime\right)^2 \right. \nonumber \\ 
		&& \left. + m_{H_2}^2 \left(\sin \alpha \cos \epsilon^\prime \sin \epsilon + \cos \alpha \sin \epsilon^\prime \right)^2 \right] - \frac{\Lambda^2}{v^2}  \frac{1}{\tan^2\beta} \,, \label{e:lam2_2hdms_u1} \\ 
		\lambda_3 &=& \frac{1}{v^2} \left[ 2 m_C^2 - \Lambda^2 + 
		m_h^2 \cos^2 \epsilon \frac{\sin 2 \alpha}{\sin 2 \beta} - m_{H_1}^2 \left(\frac{1}{2}\cos^2 \epsilon + \frac{3 }{4}\cos 2\epsilon^\prime - \frac{1}{4}\cos 2\epsilon \cos 2\epsilon^\prime \right) \frac{\sin 2 \alpha}{\sin 2 \beta}\right. \nonumber  \\ 
		&& - \left. m_{H_2}^2 \left(\frac{1}{2}\cos^2 \epsilon - \frac{3 }{4}\cos 2\epsilon^\prime + \frac{1}{4}\cos 2\epsilon \cos 2\epsilon^\prime \right) \frac{ \sin 2 \alpha}{\sin 2\beta} \nonumber \right. \\
		&& \left. +(m_{H_1}^2 - m_{H_2}^2) \sin \epsilon \sin 2\epsilon^\prime \frac{ \cos 2 \alpha}{\sin 2\beta}\right] \,, \label{e:lam3_2hdms_u1}\\ 
		\lambda_4 &=& \frac{2}{v^2} \left(\Lambda^2 -  m_C^2\right) \,, \label{e:lam4_2hdms_u1} 
	\end{eqnarray}
	\label{e:lamb_mass_2hdms_u1}
\end{subequations}
where we have replaced $\Lambda_1^2 + \Lambda_2^2 = \Lambda^2$, which carries implicitly the dependence on $\delta$ through \Eqn{e:L1L2}. For the sake of completeness,
we also express the quartic couplings $\lambda_{1s}\,, \lambda_{2s}\,,$ and  $\lambda_s$, which are specific to this
singlet-extended scenario, in terms of the above-mentioned physical parameters: 
\begin{subequations}
	\begin{eqnarray}
		\lambda_{1s} &=& \frac{\Lambda^2}{v_s^2}  \sin^2\beta + \frac{1}{2 \sqrt{2} v v_s \cos \beta} \left[\cos \alpha  \sin 2 \epsilon\left\{ 2 m_h^2 - m_{H_1}^2 - m_{H_2}^2 + (m_{H_1}^2 - m_{H_2}^2) \cos 2 \epsilon^\prime\right\} \right. \nonumber \\ 
		&& \left. + (m_{H_2}^2 - m_{H_1}^2) \cos\epsilon \sin\alpha \sin 2 \epsilon^\prime\right] \,,  \\
		\lambda_{2s} &=& \frac{\Lambda^2}{v_s^2}  \cos^2\beta + \frac{1}{2 \sqrt{2} v v_s \sin \beta} \left[\sin \alpha  \sin 2 \epsilon\left\{2 m_h^2 - m_{H_1}^2 - m_{H_2}^2 + (m_{H_1}^2 - m_{H_2}^2) \cos 2 \epsilon^\prime \right\} \right. \nonumber \\ 
		&& \left. + (m_{H_1}^2 - m_{H_2}^2) \cos\epsilon \cos\alpha \sin 2 \epsilon^\prime\right] \,,  \\
		\lambda_s &=& -\frac{v^2}{4 v_s^4} \Lambda_1^2 \sin^2 2\beta + \frac{2}{v_s^2}\left[
			3 v_s \mu_s + m_h^2 \sin^2\epsilon + 
			\cos^2\epsilon \left( m_{H_2}^2 \cos^2\epsilon^\prime + m_{H_1}^2 \sin^2\epsilon^\prime \right)\right] \,.		
	\end{eqnarray}
	\label{e:singlet_quartics_u1}
\end{subequations}
\subsection{Alignment limit in the $Z_3$-Symmetric 2HDM with a complex singlet}

The condition for the physical CP-even state $h$ to have exact SM-like couplings 
remains the same as we previously had in the $Z_2$-symmetric singlet-extended 2HDM.  
This is because the rotation matrix ${\mathscr O}_\alpha$ (defined in \Eqn{e:oalpha_2hdms}) has the same structure in this case. 
Therefore, the Higgs {\it alignment} condition in this theory also reads as,
\begin{eqnarray}
	\cos(\alpha - \beta) = 1 \quad {\rm and,} \quad \epsilon = 0 \,. \label{e:alignment1_2hdms_u1} 
\end{eqnarray}
However, to obtain the 2HDM with softly-broken $U(1)$-symmetry,
we need to ensure that the doublet fields do not mix with the singlet ones.
This can be achieved by requiring
\begin{eqnarray}
	\epsilon =0 \,, \qquad
	\epsilon^\prime = 0 \,, \qquad
	\delta = 0 \,.
\end{eqnarray}
One can easily verify that in the 2HDM limit defined above, the expressions
in \Eqn{e:lamb_mass_2hdms_u1} resemble the corresponding ones in
\Eqn{e:lamb_mass_u12hdm} if $H_1$ and $A_1$ are identified with the 
heavy CP-even Higgs eigenstate $H$ and CP-odd eignestate $A$ respectively.

\section{Parametrizing the standard contribution to nonstandard masses}
\label{s:param}
Now that we have established the notion of nonstandard sources of masses in the 2HDM context, we can now proceed to parametrize the different types of
contribution to nonstandard scalar masses. To this end, we define the following dimensionless quantities for the
pure 2HDM scenario:\footnote{
Note that for 2HDM with a softly-broken $U(1)$ symmetry, we have $\Lambda^2 = m_A^2$
which implies $f_A=0$.
}
\begin{subequations}
\begin{eqnarray}
		f_H = \frac{m_H^2 - \Lambda^2}{m_H^2} \,, \label{e:fH}\\
			f_A = \frac{m_A^2 - \Lambda^2}{m_A^2} \,, \label{e:fA}\\
	f_C = \frac{m_C^2 - \Lambda^2}{m_C^2} \,, \label{e:fC}
\end{eqnarray}
\label{e:nonstandrad_frac}
\end{subequations}
where $\Lambda^2$ is defined in \Eqn{e:Lamdef} and should be understood as a
notational shorthand for the nonstandard contribution to the masses, that is
not related to the electroweak VEV. 
Quite evidently, these dimensionless quantities capture the fraction of nonstandard squared masses $(m_H^2\,, m_A^2\,, m_C^2)$,  that arise from the electroweak VEV.\footnote{Strictly speaking, this interpretation of $f_X$ is more reliable when $m_X^2 \gg v^2$. In the 2HDM, the sign of $\Lambda^2$ is not fixed \textit{a priori}. Consequently, the magnitude of $f_X$ can exceed unity by a significant amount. However, as we will discuss in Sec.~\ref{s:results}, such a scenario arises only when the nonstandard scalar masses lie near the electroweak scale.}
Now the key question is whether these fractions can be constrained through phenomenological considerations.
In the followings, we will explore different possibilities to find an affirmative answer.

To this end, we first provide a list of certain trilinear Higgs self-couplings that are directly connected to
the different `fractions' defined in \Eqn{e:nonstandrad_frac}. To establish the notations, let us write
the relevant part of the interaction Lagrangian as follows:
\begin{eqnarray}
	{\mathscr L}_{\rm int} &=& \lambda_{hH^+H^-} hH^+H^- +\lambda_{HH^+H^-} HH^+H^- +\frac{1}{2} \lambda_{hHH} hHH + \frac{1}{2} \lambda_{hAA} hAA \nonumber \\
	&& +\frac{1}{2} \lambda_{HHH} HHH +\frac{1}{2} \lambda_{HAA} HAA  \,.
	\label{e:lamdef}
\end{eqnarray}
In the {\em alignment limit} of 2HDM, the expressions for the above couplings
assume the following form (the general expressions are given in appendix~\ref{app:gen_coup}):
\begin{subequations}
	\label{e:triself}
	\begin{eqnarray}
		\lambda_{hH^+H^-} &=& -\frac{2}{v}\left(m_C^2 -\Lambda^2 -\frac{m_h^2}{2} \right)
		\equiv -\frac{2m_C^2}{v}\left(f_C +\frac{m_h^2}{2m_C^2} \right) \,, \label{e:hhphm}\\
		\lambda_{hHH} &=& -\frac{2}{v}\left(m_H^2 -\Lambda^2 +\frac{m_h^2}{2} \right)
		\equiv -\frac{2m_H^2}{v}\left(f_H +\frac{m_h^2}{2m_H^2} \right) \,, \\	
		\lambda_{hAA} &=& -\frac{2}{v}\left(m_A^2 -\Lambda^2 +\frac{m_h^2}{2} \right)
		\equiv -\frac{2m_A^2}{v}\left(f_A +\frac{m_h^2}{2m_A^2} \right) \,, \\
		\lambda_{HH^+H^-} = \lambda_{HHH} = \lambda_{HAA} &=& -\frac{2}{v}\left(m_H^2 -\Lambda^2\right)\cot 2\beta
		\equiv -\frac{2m_H^2}{v}f_H \cot 2\beta   \label{e:H1hphm} \,.    
	\end{eqnarray}
\end{subequations}
These interactions can, in principle, be constrained at colliders through various experimental searches. Such experimental limits can then be translated into bounds on the fractions defined in \Eqn{e:nonstandrad_frac}. Among the couplings appearing in \Eqn{e:triself}, preliminary bounds can already be placed on $\lambda_{hH^+H^-}$ and $\lambda_{HH^+H^-}$ from measurements of $\mu_{\gamma\gamma}$~\cite{ATLAS:2022tnm} and from searches for $pp \to H \to \gamma\gamma$~\cite{ATLAS:2021uiz, CMS:2024nht}, respectively.
The coupling $\lambda_{hH^+H^-}$ affects the signal strength of the SM-like Higgs boson ($h$) in the $\gamma\gamma$ decay channel through additional loop contributions from the charged scalar. Consequently, current limits on $\mu_{\gamma\gamma}$ can be used to constrain $\lambda_{hH^+H^-}$, which can then be interpreted as a bound on $f_C$ using \Eqn{e:hhphm}.
Similarly, $\lambda_{HH^+H^-}$ influences the $pp \to H \to \gamma\gamma$ production cross-section. Thus, direct search limits on $\sigma(pp \to H \to \gamma\gamma)$ can be used to constrain $\lambda_{HH^+H^-}$, which in turn can be translated into a bound on $f_H$ via \Eqn{e:H1hphm}.


But before delving into these processes, we first examine the theory constraints in the next section.

\section{Constraints from unitarity, stability and the $T$-parameter}
\label{sec:theorybound}
The main theoretical constraints come from  perturbative unitarity and stability
of the scalar potential. The necessary and
sufficient conditions for the potential to be bounded-from-below (BFB) are given as~\cite{Deshpande:1977rw,Klimenko:1984qx},
\begin{eqnarray}
\lambda_1 \geq 0 \,, \quad \lambda_2 \geq 0 \,, \quad
\lambda_3 \geq - \sqrt{\lambda_{1} \lambda_{2}} \,, \quad \lambda_3 + \lambda_4 - |\lambda_5| \geq \sqrt{\lambda_{1} \lambda_{2}} \,.
\label{e:BFB}
\end{eqnarray}
The conditions of tree-unitarity for a 2HDM with softly-broken $Z_2$-symmetry essentially translate into bounds on the following eigenvalues of the 
$S$-matrix~\cite{Maalampi:1991fb,Kanemura:1993hm,Akeroyd:2000wc,Horejsi:2005da}:
\begin{subequations}
	\begin{eqnarray}
	 a_1^\pm &=& \frac{1}{2} \left(3 \lambda_1 + 3 \lambda_2 \pm \sqrt{
			9 (\lambda_1 - \lambda_2)^2 + 4 (2 \lambda_3 + \lambda_4)^2}\right) \,, \\
	 a_2^\pm &=& \frac{1}{2} \left(\lambda_1 + \lambda_2 \pm 
			\sqrt{(\lambda_1 - \lambda_2)^2 + 4 \lambda_4^2}\right) \,, \\  
	 a_3^\pm &=& \frac{1}{2} \left(\lambda_1 + \lambda_2 \pm 
			\sqrt{(\lambda_1 - \lambda_2)^2 + 4 \lambda_5^2}\right) \,, \\ 
		b_1^\pm &=& \lambda_3 + 2 \lambda_4 \pm 3 \lambda_5\,, \\
		b_2^\pm &=& \lambda_3 \pm \lambda_5 \,, \\
		b_5^\pm &=& \lambda_3 \pm \lambda_4 \,.
	\end{eqnarray}
\label{e:Uni}
\end{subequations}
The requirement of tree-level unitarity imposes an upper bound on each of the eigenvalues as follows:
\begin{eqnarray}
	|a_i^\pm| \,, \, |b_i^\pm| \leq 8 \pi \,.
\end{eqnarray}
We will first analyze the 2HDM parameter space in light of the above unitarity and BFB constraints. In our analysis, we will always assume the alignment limit to hold so that
the constraints from the Higgs signal strenghts, for the tree level decays of the
SM-like Higgs boson, are automatically satisfied.
We begin by examining some intriguing features of the BFB and unitarity conditions
presented above. The first two BFB constraints in \Eqn{e:BFB} imply
\begin{eqnarray}
	\lambda_{1} + \lambda_{2} \geq 0 \,.
\end{eqnarray}
This in conjuction with the unitarity condition $|a_1^\pm| \leq 8 \pi$ 
should lead to~\cite{Das:2015qva}
\begin{subequations}
\begin{eqnarray}
	0 \leq \lambda_{1} + \lambda_{2} \leq \frac{16 \pi}{3} \,.
\end{eqnarray}
Using \Eqn{e:lamb_mass_2hdm} the above condition can be expressed in terms of
the physical parameters as follows:
\begin{eqnarray}
&& 0 \leq \left(m_H^2 - \Lambda^2 \right)\left(\tan^2 \beta + \cot^2 \beta \right) + 2 m_h^2 \leq \frac{16 \pi v^2}{3} \,, \\
\Rightarrow && 0 \leq f_H m_H^2 \left(\tan^2 \beta + \cot^2 \beta \right) + 2 m_h^2 \leq \frac{16 \pi v^2}{3} \,. \label{e:fHuni}
\end{eqnarray}
\label{e:uni_bfb1}
\end{subequations}
Furthermore, one can also write:
\begin{subequations}
	\begin{eqnarray} 
	&&	|b_1^- - b_1^+| \equiv 6 |\lambda_5| \leq 16 \pi  \,, \\
	&&	|b_5^- - b_2^+| \equiv 6 |\lambda_4 + \lambda_5| \leq 16 \pi   \,,
	\end{eqnarray}
\label{e:uni_bfb2}
\end{subequations}
which can be converted into the following conditions:
\begin{subequations}
	\begin{eqnarray} 
 |m_A^2 - \Lambda^2| < \dfrac{8  \pi v^2}{3} && \Rightarrow \, |f_A| m_A^2 < \dfrac{8  \pi v^2}{3} \,, \label{e:fAui} \\
 |m_C^2 - \Lambda^2| < \dfrac{8  \pi v^2}{3}  && \Rightarrow \, |f_C| m_C^2 < \dfrac{8 \pi v^2 }{3} \label{e:fCuni}  \,.
	\end{eqnarray}
	\label{e:fCfAuni}
\end{subequations}
An important observation from \Eqs{e:uni_bfb1}{e:fCfAuni} is that when $\Lambda^2 = 0$, {\it i.e.}, $f_H = f_C = f_A = 1$, the unitarity and BFB constraints translate directly into upper bounds on the nonstandard scalar masses. The upper limit on $m_H$, in particular, depends on $\tan\beta$, as seen from \Eqn{e:uni_bfb1}. Consequently, if a nonstandard scalar is observed in the future with a mass exceeding this limit, it would indicate that a portion of its mass must originate from the soft symmetry-breaking term.
Furthermore, in the limit $m_X^2 \gg v^2$ ($X \equiv H, A, C$), \Eqs{e:fHuni}{e:fCfAuni} require that $f_X \approx 0$. This implies that for the nonstandard scalars to be much heavier than the electroweak scale, their masses must arise almost entirely from the soft-breaking parameter, which -- as demonstrated earlier -- encapsulates the nonstandard contribution to the scalar mass spectrum.


In passing, we note that the splitting between nonstandard scalar masses is also
constrained by the new physics contribution to the electroweak $T$-parameter, whose
expression in the alignment limit is given by~\cite{He:2001tp,Grimus:2007if, Grimus:2008nb}
\begin{eqnarray}
{\Delta T}_{\rm NP} = {1 \over 16\pi \sin^2 \theta_w M_W^2} \Big[{\mathscr F}(m_C^2,m_H^2) +
{\mathscr F}(m_C^2, m_A^2) - {\mathscr F}(m_H^2,m_A^2) \Big] \,,
\label{e:T}
\end{eqnarray}
with
\begin{eqnarray}
{\mathscr F}(x,y) = {x+y \over 2} - {xy \over x-y} \, \ln\left(\frac{x}{y} \right) \,.
\label{eq:Ffunc}
\end{eqnarray}
In \Eqn{e:T}, $M_W$ represents the mass of the $W$-boson and $\theta_w$ is the weak mixing angle.
The current limit on ${\Delta T}_{\rm NP}$ is given by~\cite{ParticleDataGroup:2024cfk},
\begin{eqnarray}
\left({\Delta T}_{\rm NP}\right)^{\rm fit} = 0.04 \pm 0.12 \,.
\end{eqnarray}
Next we will investigate whether additional constraints can arise from current LHC data.

\section{Implications from di-photon searches}
\label{s:diphoton}
The lightest CP-even Higgs boson ($h$) exhibits SM-like couplings in the alignment limit. However, its loop-induced decay channel into two photons receives additional nonstandard contributions from the charged-Higgs loop. Consequently, the charged-Higgs coupling to $h$, as given in Eq.~(\ref{e:hhphm}), and hence the fraction $f_C$, plays an important role in this decay process. To illustrate this, the expression for the Higgs signal strength in the diphoton channel, $\mu_{\gamma\gamma}$, can be written as follows~\cite{Bhattacharyya:2014oka}:

\begin{eqnarray}
	\mu_{\gamma \gamma} \equiv \left|\kappa_\gamma^h \right|^2 = \dfrac{\Gamma_{\rm 2HDM}(h \to \gamma \gamma)}{\Gamma_{\rm SM}(h \to \gamma \gamma)} = \dfrac{\left| F_W(\tau_W^h) + \frac{4}{3} F_t (\tau_t^h) + \kappa_C^h F_C (\tau_C^h) \right|^2}{\left\lvert F_W(\tau_W^h) + \frac{4}{3} F_t (\tau_t^h) \right\rvert^2}\,, 
\end{eqnarray}
where $\tau_x^h = (2m_x/m_h)^2$ and the functions are given by~\cite{Gunion:1989we}
\begin{subequations}
\begin{eqnarray}
	F_W(x) &=& 2+3x+3x(2-x)f(x) \, , \\
	F_t(x) &=& -2x\left\{1+(1-x)f(x) \right\} \, , \\
	F_C (x) &=&-x \left\{1-xf(x)\right\} \,, \\
	{\rm with,}~~ f(x) &=& \begin{cases}
	\left[\sin^{-1}\left(\sqrt{\frac{1}{x}} \right) \right]^2 \,, & \text{for } x \ge 1 \,, \\
	-\frac{1}{4}\left[\ln\left(\frac{1+\sqrt{1-x}}{1-\sqrt{1-x}}\right)-i\pi \right]^2 \,, & \text{for } x < 1 \,.
	\end{cases}
\end{eqnarray}
\label{e:gamgam_loop}
\end{subequations}
%
The coefficient $\kappa_C^h$ is derived from $\lambda_{hH^+H^-}$, defined in Eq.~(\ref{e:hhphm}),  as follows:
\begin{eqnarray}
	\kappa_C^h &=& \dfrac{v}{2 m_C^2} \lambda_{hH^+H^-} = - \left( f_C + \dfrac{m_h^2}{2 m_C^2}\right) \, , 
	\label{e:kch}
\end{eqnarray}

It is interesting to note that the signal strength $\mu_{\gamma \gamma}$, in the alignment limit, is only a function of $\left( m_C^2 \, , f_C \right)$ and considering the 
current experimental data~\cite{ATLAS:2022tnm}, 
\begin{eqnarray}
	\mu_{\gamma \gamma} = 1.04^{+0.10}_{-0.09} \, ,
	\label{e:mugam_LHC}
\end{eqnarray}
one can restrict the parameter space in the $\left( m_C \, , f_C \right)$ plane.
One can also utilize the anticipated future sensitivities at the HL-LHC and the ILC~\cite{Fujii:2015jha} to project the expected constraints in the $f_C$–$m_C$ plane. Assuming that future Higgs signal-strength measurements will continue to be consistent with the Standard Model (SM) expectations, we adopt a projected uncertainty of $1.9\%$, as reported in Ref.~\cite{deBlas:2019rxi}:
\begin{eqnarray}
\kappa_\gamma^h = 1 \pm 0.019 \,.
\label{e:kgfuture}
\end{eqnarray}
%


On the other hand, the heavy CP-even Higgs boson, $H$, also couples to the charged Higgs and can decay into a diphoton final state through the charged particles in the loop. The corresponding decay width in the 2HDM is given by
\begin{eqnarray}
\Gamma(H \to \gamma \gamma)_{\rm 2HDM} 
= \frac{\alpha_e^2 g^2}{2^{10} \pi^3}
\frac{m_H^3}{M_W^2}
\left\lvert 
\kappa_W^H F_W(\tau_W^H)
+ \frac{4}{3} \kappa_t^H F_t(\tau_t^H)
+ \kappa_C^H F_C(\tau_C^H)
\right\rvert^2 \,.
\label{e:H2gg}
\end{eqnarray}
Here, $g$ denotes the $SU(2)_L$ gauge coupling and $\alpha_e$ is the fine-structure constant. The loop functions are identical to those defined in Eq.~(\ref{e:gamgam_loop}), with the identification $\tau_x^H \equiv (2m_x/m_H)^2$.

The coupling modifiers $\kappa_W^H$ and $\kappa_t^H$ are defined as the ratios of the heavy Higgs couplings to $W$ and $t$ relative to their SM counterparts, {\it i.e.},
\begin{eqnarray}
\kappa_x^H = \dfrac{g_{Hxx}}{g_{hxx}^{\rm SM}} \,, 
\quad \{x = W,\, t\} \,.
\label{e:kH}
\end{eqnarray}
In the alignment limit, however, the trilinear heavy Higgs couplings to gauge bosons    of the form $ HV^\mu V_\mu$ $(V \equiv W^\pm,\, Z)$ vanish. Consequently, the coupling modifier $\kappa_W^H$ becomes zero, implying that only the top-quark loop and the charged Higgs loop contribute to the $H\to \gamma\gamma$ decay width.

As a result, the dominant production mechanism for the heavy Higgs boson is gluon-gluon fusion, where the cross section is modified only through the altered top coupling. Hence, the  cross section for the process $pp \to H \to \gamma\gamma$ can be expressed as\footnote{We have also considered a similar bound for $pp \to A \to \gamma \gamma$ in an analagous way, keeping in mind that only the top loop should contribute to the decay.}
\begin{eqnarray}
\sigma(pp \to H \to \gamma\gamma) =
\sigma_{\rm prod} \cdot {\rm BR}^H_{\gamma \gamma} 
= (\kappa_t^H)^2 \, \sigma_{ggH}^{\rm SM} \times {\rm BR}(H \to \gamma \gamma)_{\rm 2HDM} \,.
\end{eqnarray}
The coefficient $\kappa_C^H$ appearing in \Eqn{e:H2gg} is defined as
\begin{eqnarray}
\kappa_C^H 
\equiv \dfrac{v}{2 m_C^2} \lambda_{HH^+H^-} 
= - \dfrac{m_H^2}{m_C^2} f_H \cot 2\beta \,,
\label{e:kappac}
\end{eqnarray}
where \Eqs{e:lamdef}{e:H1hphm} have been used.
Currently, the ATLAS~\cite{ATLAS:2021uiz} and CMS~\cite{CMS:2024nht} collaborations have reported upper bounds on 
$\sigma(pp \to H \to \gamma \gamma)$ for a range of heavy Higgs masses $(m_H)$. 
Therefore, while the SM Higgs data constrain the mass fraction $f_C$, 
the experimental limits from heavy Higgs searches in the diphoton channel 
allow us to place independent bounds on the nonstandard scalar mass fraction $f_H$. 
In the following section, we present a detailed discussion of our results and implications.

\section{Results}
\label{s:results}
In Sec.~\ref{s:param} we introduced a convenient parametrization that separates the electroweak-VEV-induced and non-electroweak contributions to the nonstandard scalar masses. The main goal of this section is to examine whether the fractions $f_X$ ($X \equiv H, A, C$), defined in \Eqn{e:nonstandrad_frac}, can be constrained or probed from TeV-scale phenomenology. For the numerical analysis, we take
\[
\{m_H,\, m_A,\, m_C,\, \log_{10}\tan\beta,\, \Lambda\}
\]
as the set of independent input parameters. Throughout the analysis we impose the 2HDM alignment limit of \Eqn{e:alignment_2hdm}; hence the angle $\alpha$ is computed internally from the chosen value of $\tan\beta$.

We generate $\mathcal{O}(10^{6})$ random points in the above parameter space, varying the BSM scalar masses and $\tan\beta$ over the ranges
\begin{subequations}
\begin{eqnarray}
m_{H,A,C} \in \left[125~\text{GeV},\, 2~\text{TeV}\right], 
\qquad 
\log_{10}(\tan\beta) \in [-1,\, 1.5] \,.
\label{e:BSMmass}
\end{eqnarray}
To generate random values of $\Lambda^2$, which encodes the effect of the soft-breaking parameter, we define the difference
\begin{eqnarray}
\Delta = m_H^2 - \Lambda^2\,,
\label{e:deldef}
\end{eqnarray}
which is tightly constrained by \Eqn{e:uni_bfb1}. To automatically satisfy this constraint and improve the acceptance rate of random points, we first sample $\Delta$ uniformly within
\begin{equation}
\Delta \in 
\left[
\frac{\Delta_{\rm min}}{\tan^{2}\beta + \cot^{2}\beta},\,
\frac{\Delta_{\rm max}}{\tan^{2}\beta + \cot^{2}\beta}
\right],
\end{equation}
where $\Delta_{\rm min} = - 2m_{h}^{2}$ and 
$\Delta_{\rm max} = \frac{16}{3}\pi v^{2} - 2m_{h}^{2}$.  
Finally, using \Eqn{e:deldef}, we compute
\begin{eqnarray}
\Lambda^{2} = m_{H}^{2} - \Delta\,.
\end{eqnarray}
\end{subequations}

\begin{figure}[htbp!]
	\centering
	\includegraphics[scale=0.3]{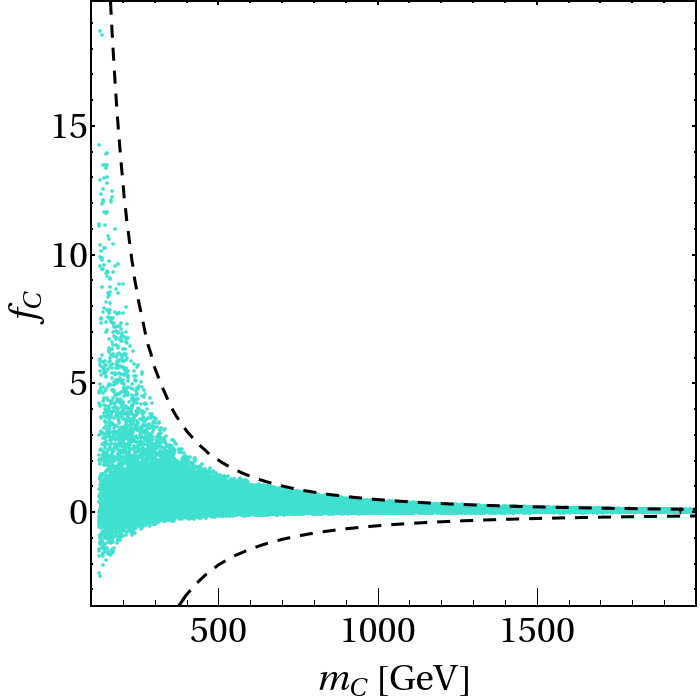}
	\includegraphics[scale=0.3]{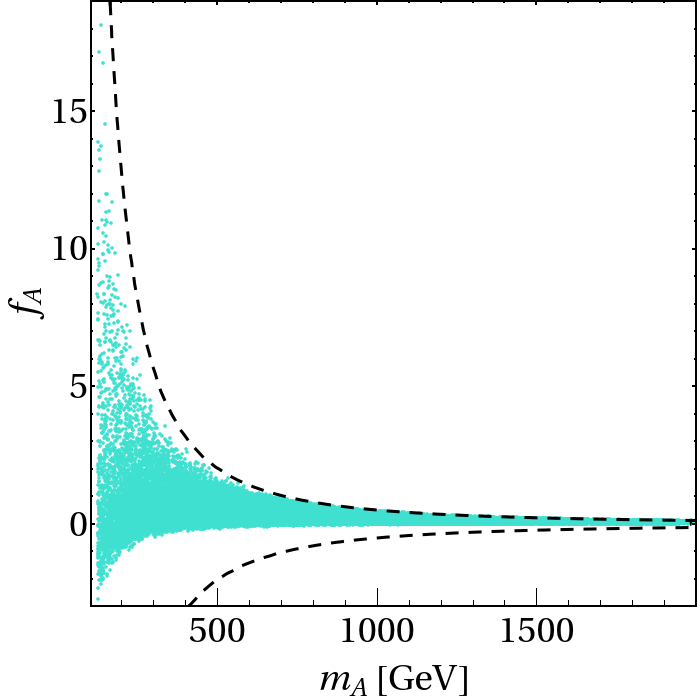} \\
	\vspace{0.5cm}
	\includegraphics[scale=0.3]{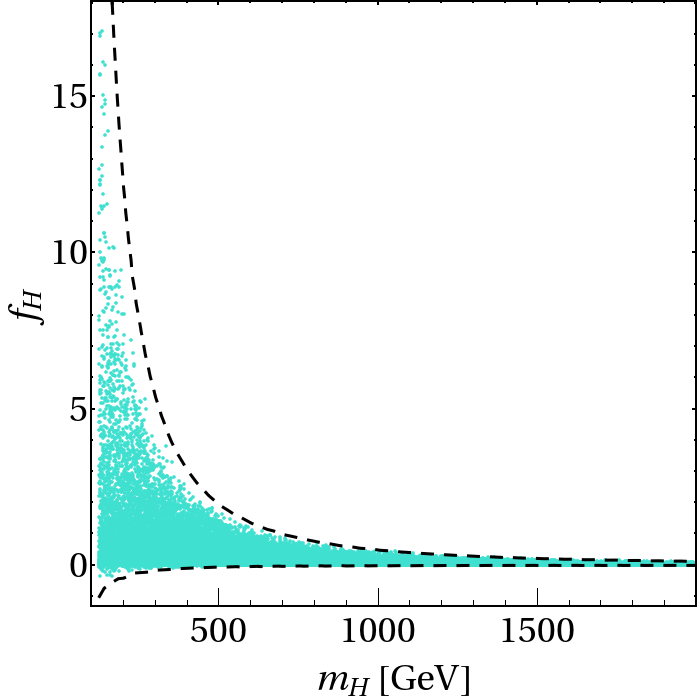}
	\includegraphics[scale=0.3]{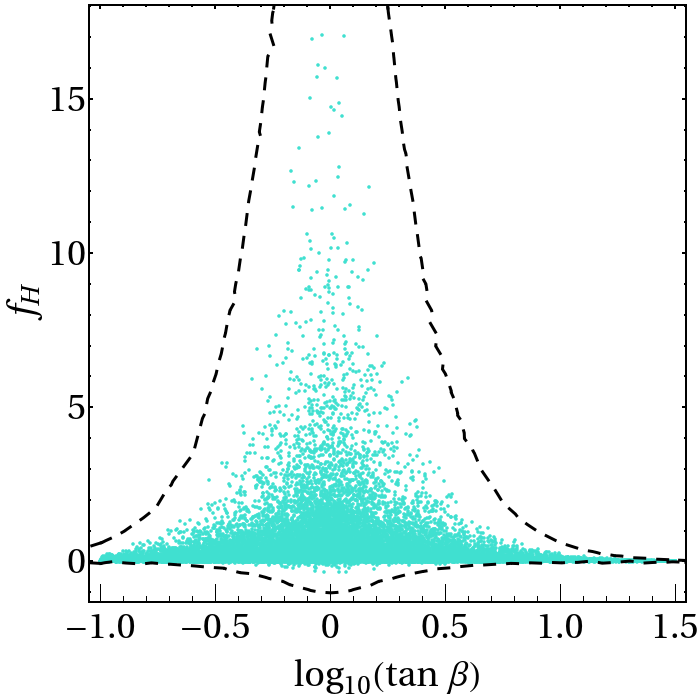}
	\caption{The scattered points indicate the allowed regions in the various parameter planes that satisfy the constraints from perturbative unitarity, boundedness from below (BFB), and the electroweak $T$-parameter.  The dashed black outer contours in the top two panels correspond to the constraints derived from \Eqn{e:fCfAuni}. Similarly, the dashed black outer contours in the bottom two panels arise from \Eqn{e:uni_bfb1}. For the bottom panels, we have chosen $\tan\beta = 1$ and $m_H = m_h$, respectively, as these yield the most relaxed constraints.
    Note that these constraints only depend on the scalar sector of the 2HDM and
	therefore are valid for all the 2HDM variants that feature a softly-broken
	$Z_2$ symmetry.}
	\label{fig:uniMvsf}
\end{figure}

As a first step, we subject the randomly generated points to the unitarity, BFB, and $T$-parameter constraints discussed in Sec.~\ref{sec:theorybound}. The points that satisfy all these constraints are shown in different parameter planes  in Fig.~\ref{fig:uniMvsf}. A notable feature of these allowed regions is that, for relatively light nonstandard masses $m_{X} \lesssim 400~\text{GeV}$, ($X \equiv H, C, A$) the corresponding $f_{X}$ values can be quite large. This indicates that, in this regime, the contribution from the electroweak VEV to the nonstandard masses can dominate over the contribution from $\Lambda^{2}$. In contrast, for $m_{X} \gtrsim 1~\text{TeV}$ we find $|f_{X}| < 0.8$, which implies that for nonstandard scalar masses above the TeV scale, a non-negligible contribution from $\Lambda^{2}$ becomes necessary. Both these observations follow directly from the upper bounds given in \Eqn{e:fCfAuni}.  

For the pseudoscalar and charged scalar masses, the theoretical constraints impose the following upper limits:
%
\begin{eqnarray}
|f_{A}| < \frac{8\pi v^{2}}{3 m_A^2}, \qquad 
|f_{C}| < \frac{8\pi v^{2}}{3 m_C^2} \,.
\label{e:fxmxAC}
\end{eqnarray}
These relations qualitatively explain the behavior of the $f_{X}$ limits both for $m_{X} \sim \order(v)$ and for the hierarchical regime $m_{X} \gg v$.
To provide a visual guideline, these constraints have been shown as the dashed black outer contours in the first two panels of Fig.~\ref{fig:uniMvsf}.
The qualitative nature of the constraints on $f_H$ arising from unitarity, BFB conditions, and the electroweak $T$-parameter can, on the other hand, be anticipated from \Eqn{e:uni_bfb1}. The corresponding allowed regions are shown by the scattered points in the bottom two panels of Fig.~\ref{fig:uniMvsf}, while the overall extent of the constraints imposed by \Eqn{e:uni_bfb1} is indicated by the black dashed outer contours in these panels.
To obtain the contours in the $f_H$-$m_H$ and $f_H$-$\log_{10}\tan\beta$ planes, we fix $\tan\beta = 1$ and $m_H = m_h$, respectively, choices that lead to the most conservative ({\it i.e.}\ least restrictive) bounds. A noteworthy feature of the fourth panel of Fig.~\ref{fig:uniMvsf} is that the constraint on $f_H$ is weakest for $\tan\beta \sim \order(1)$, as anticipated from \Eqn{e:uni_bfb1}.

\begin{figure}[htb!]
	\centering
	\includegraphics[scale=0.22]{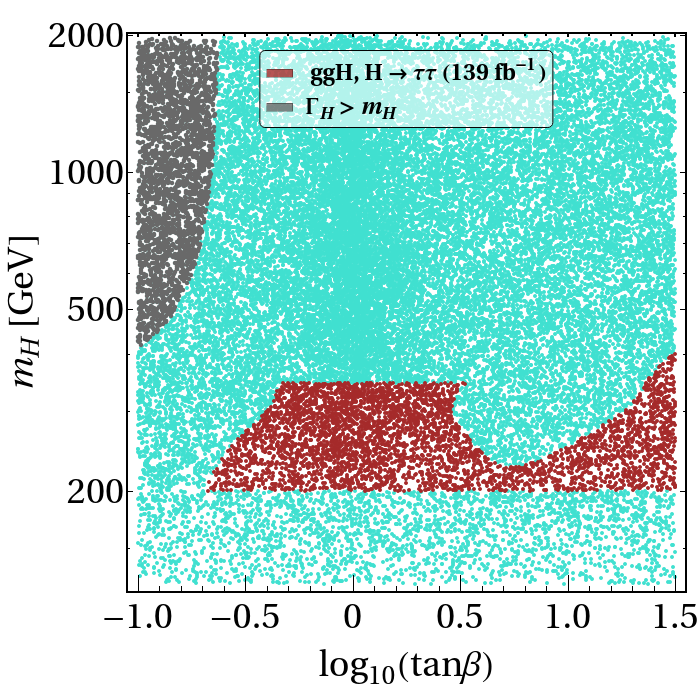}
	\includegraphics[scale=0.22]{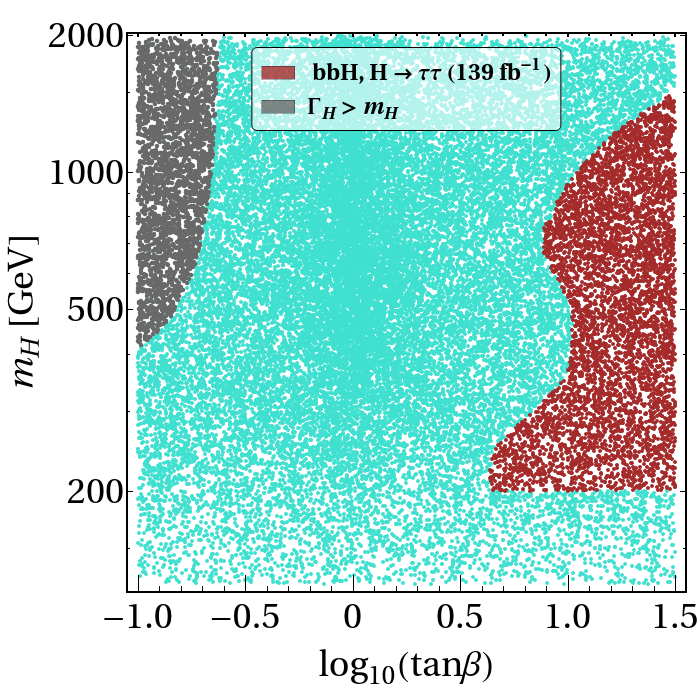}
	\includegraphics[scale=0.22]{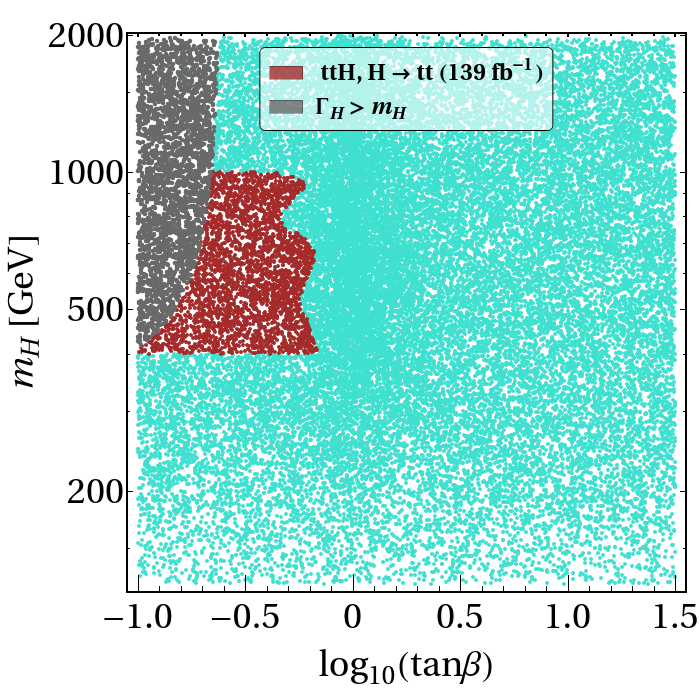} \\
	\includegraphics[scale=0.22]{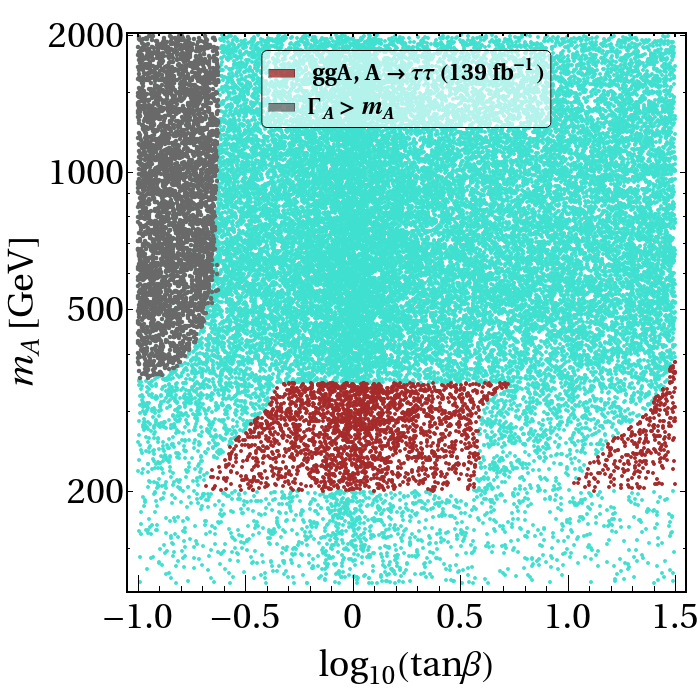}
	\includegraphics[scale=0.22]{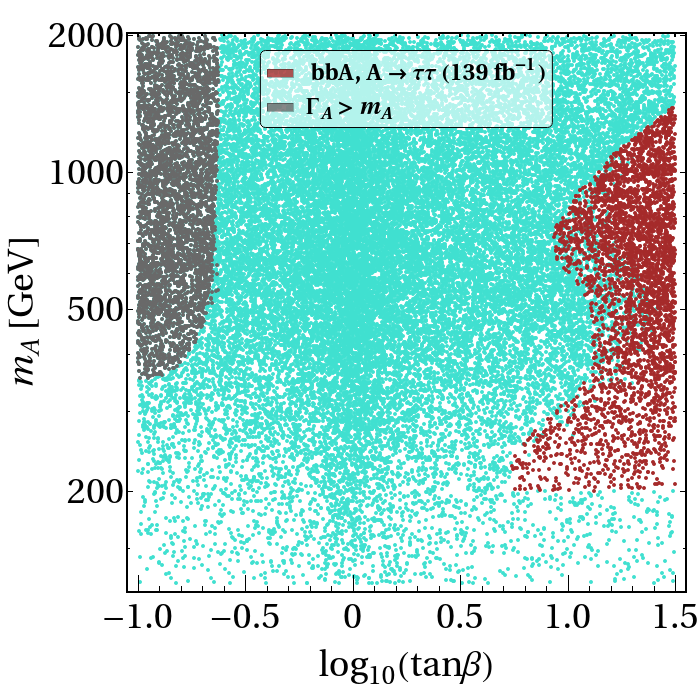}
	\includegraphics[scale=0.22]{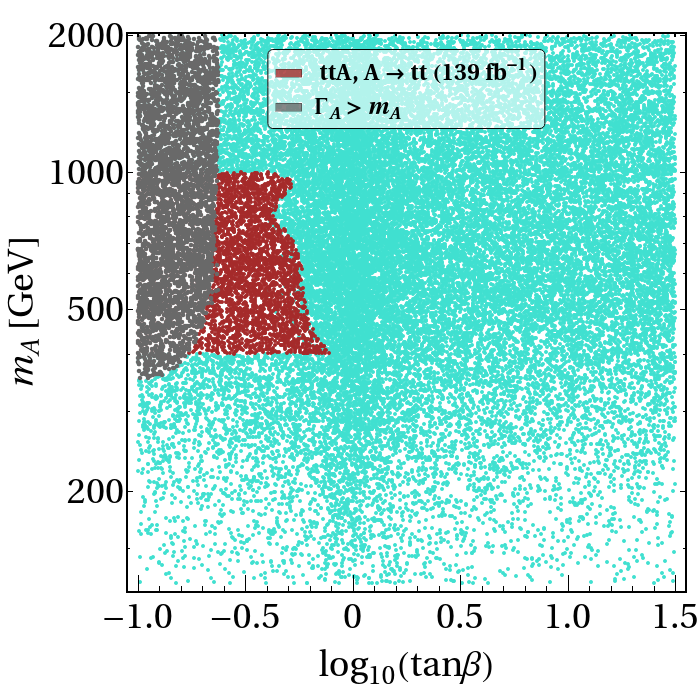} 
	\caption{\it The allowed parameter space for the CP-even (H) and CP-odd (A) neutral scalars is shown in the upper and lower panels, respectively. Red points in the left (middle) column are excluded by $\tau\tau$ searches from gluon-fusion (bottom-associated) production, while those in the right column are excluded by $t\bar{t}$ bounds from top-associated production. These bounds from the direct searches are derived assuming a type-II Yukawa structure. Gray points in all panels are disfavored by the requirement $\Gamma_{H(A)} < m_{H(A)}$, where $\Gamma_{H(A)}$ is the total decay width of H(A). The teal background points satisfy all the constraints
	in Fig.~\ref{fig:uniMvsf}.}
	\label{f:Hsearch}
\end{figure}

Before turning to the constraints from $h,H \to \gamma\gamma$ and presenting them in the $f_X$ versus $m_X$ ($X \equiv H, C$) planes -- which constitute the main agenda of this section -- it is important to recall that additional bounds arise from direct searches such as $pp \to H/A \to \tau^+\tau^-$, $pp \to H/A \to \bar{t}t$, $pp \to H^+ \to \tau^+\nu$, and $pp \to H^+ \to t\bar{b}$.\footnote{
Because of the alignment limit, additional constraints will not arise from direct search channels such as $pp \to H \to W^+W^-$, $pp \to H \to ZZ$, $pp \to A \to Zh$ etc.
} These channels can exclude further portions of the parameter space from the allowed regions in Fig.~\ref{fig:uniMvsf}.
However, the constraints from these processes are not directly sensitive to $f_X$, and consequently the corresponding exclusions are not expected to be cleanly separated in the $f_X$ vs. $m_X$ planes. Nevertheless, these bounds remain phenomenologically important. To illustrate their impact more transparently, we therefore present the corresponding constraints in the $\log_{10}\tan\beta$ vs. $m_X$ ($X \equiv H, C, A$) planes, where the excluded regions are expected to be well segregated from the allowed ones.
Subsequently, when discussing the final constraints in the $f_X$ versus $m_X$ ($X \equiv H, C$) planes, we retain only those parameter points that satisfy the combined constraints from unitarity, BFB conditions, the electroweak $T$-parameter, and the direct search limits from the two-body fermionic channels.

Keeping these considerations in mind, we begin with searches for neutral scalars. In a type-II 2HDM, both the CP-even scalar $H$ and the CP-odd scalar $A$ can decay into $\tau^{+}\tau^{-}$ and $t\bar{t}$ final states with sizable branching fractions for specific ranges of $\tan\beta$. The ATLAS collaboration has searched for such neutral Higgs bosons in the $\tau^{+}\tau^{-}$ channel\cite{ATLAS:2020zms} over the mass range $0.2$-$2.5~\mathrm{TeV}$, considering two production mechanisms: gluon-gluon fusion and $b$-associated production. The absence of any statistically significant excess above the SM background allows ATLAS to place upper limits on the corresponding production cross sections.
These limits can be translated into constraints in the $\log_{10}\tan\beta$-$m_{H,A}$ planes, as shown in the first two panels of Figs.~\ref{f:Hsearch}. 
The excluded regions are indicated by red points. In addition, the third panels of Figs.~\ref{f:Hsearch} 
display the excluded regions derived from neutral Higgs searches in the $t\bar{t}$ channel,\cite{ATLAS:2022rws} performed in the mass range $400~\mathrm{GeV}$ to $1~\mathrm{TeV}$, assuming production via the top-associated channel.
The qualitative features of the red excluded regions in Fig.~\ref{f:Hsearch} can be understood by noting that the coupling modifiers $\kappa^{H, A}_{t}$ and $\kappa^{H, A}_{\tau,b}$, defined analogously to \Eqn{e:kH}, scale as $\cot\beta$ and $\tan\beta$, respectively. 
Finally, the regions highlighted by gray points in Figs.~\ref{f:Hsearch} 
 are disfavored by the requirement $\Gamma_{H,A} < m_{H,A}$, where $\Gamma_H$ ($\Gamma_A$) denotes the total decay width of $H$ ($A$).\footnote{This condition may be viewed as a consequence of tree-level unitarity\cite{Bandyopadhyay:2018cwu} in $t\bar{t}\to t\bar{t}$ scattering and ensures that $H$ and $A$ admit a consistent particle interpretation.}

\begin{figure}[htb!]
	\centering
	\includegraphics[scale=0.3]{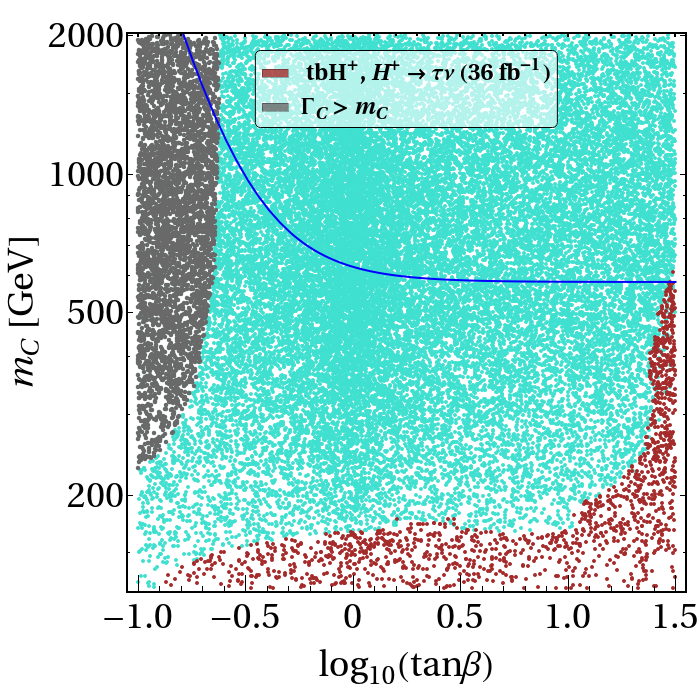}
	\includegraphics[scale=0.3]{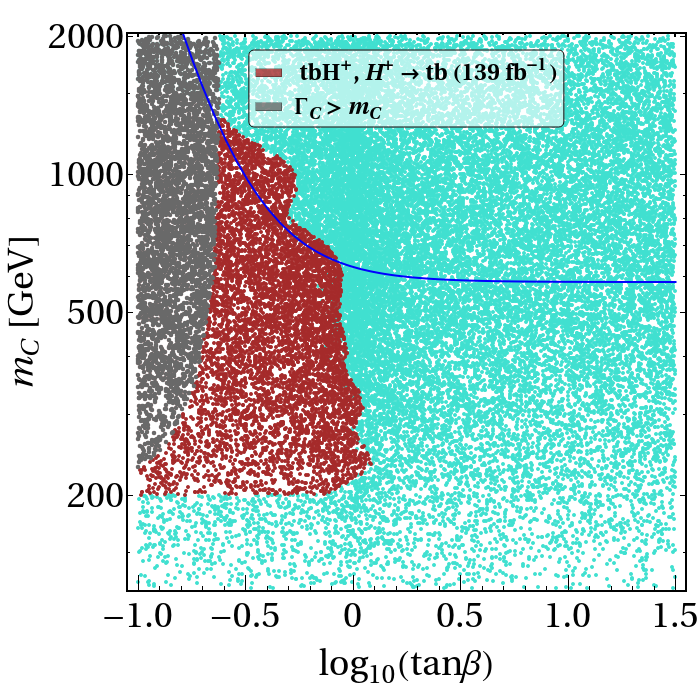}
	\caption{In the left (right) panel, the  red points are excluded by charged Higgs searches in the $\tau\nu$ ($t\bar{b}$) channel\cite{ATLAS:2018gfm,ATLAS:2021upq}. In both these cases, the charged Higgs is produced via $pp\to H^{+} t \bar{b}$ mode. In both panels, the regions indicated by  gray points are disfavored by the requirement $\Gamma_C < m_C$, where $\Gamma_C$ denotes the total decay width of the charged scalar. The exclusion contour arising from the $b \to s\gamma$ constraint\cite{Misiak:2017bgg,Atkinson:2021eox} is shown as a  blue solid line. The entire range of $\tan\beta$ is excluded due to this bound for $m_{C} < 580$~GeV. The bounds from the direct searches and $b \to s\gamma$ are derived assuming a type-II Yukawa structure.}
	\label{f:Csearch}
\end{figure}

In the left and right panels of Fig.~\ref{f:Csearch}, we display the regions constrained by the
$H^+ \to \tau^+ \nu$ and $H^+ \to t\bar{b}$ search channels~\cite{ATLAS:2018gfm,ATLAS:2021upq},
respectively. The regions excluded by these searches are indicated by the red scattered points.
The qualitative features of these exclusions can be understood from the structure of the charged
Higgs Yukawa interactions. In a type-II two-Higgs-doublet model, these interactions are given by 
\begin{eqnarray}
\mathscr{L}_C^{\rm type-II}
= \frac{\sqrt{2}}{v}\, H^+ \,
\left\{\bar{t}\left( m_t \cot\beta\, P_L + m_b \tan\beta\, P_R \right) b 
+\bar{\nu}_\tau\, m_\tau \tan\beta\, P_R\, \tau \right\}
\; + \; \mathrm{h.c.}
\end{eqnarray}
These interactions also account for the constraint arising from $b \to s \gamma$, shown in
Fig.~\ref{f:Csearch} by the  blue solid line, which we take from Ref.~\cite{Misiak:2017bgg}.
Finally, the gray points in Fig.~\ref{f:Csearch} correspond to regions disfavored by the
requirement $\Gamma_C < m_C$, where $\Gamma_C$ is the total decay width of the charged Higgs boson.
This width depends sensitively on the value of $\tan\beta$.

\begin{figure}[htb!]
	\centering
	\includegraphics[scale=0.28]{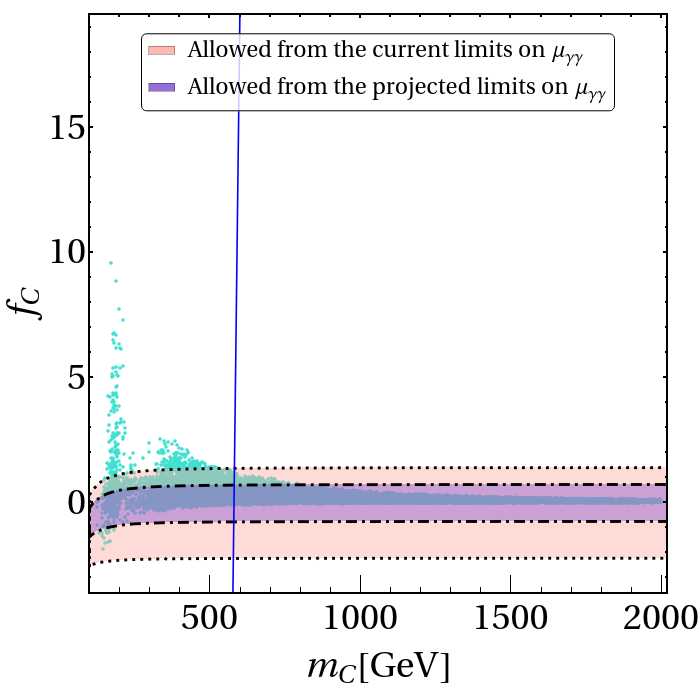} ~~~
	\includegraphics[scale=0.28]{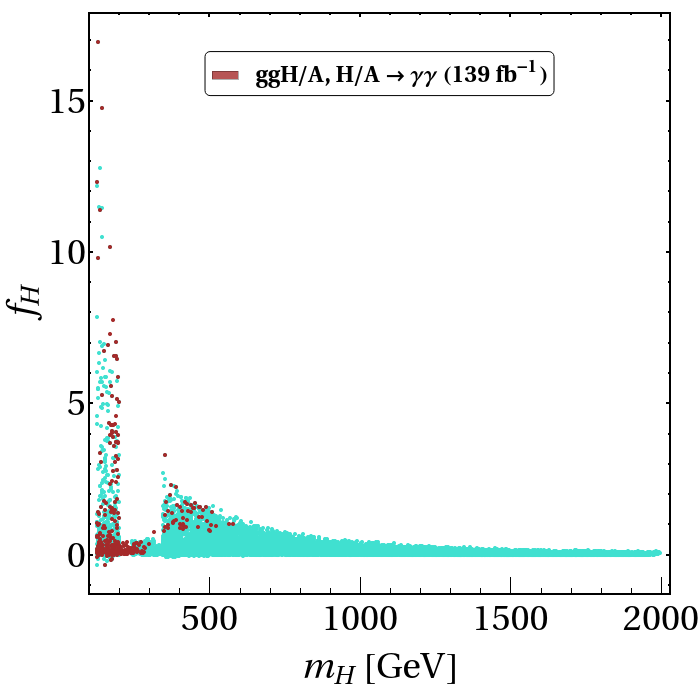}
	\caption{In the left (right) panel, the  teal scattered points satisfy the combined constraints from unitarity, BFB conditions, the $T$-parameter and direct LHC search limit. In the left panel light pink (purple)  colored band represents allowed region which is consistent with current (future) $\mu_{\gamma\gamma}$ measurements at 95\% C.L. In the right panel, the  red scattered points are disallowed from direct search limits arising from $p p \to H/A \to \gamma \gamma$ process. We note that these constraints are specific to the type-II Yukawa structure of the 2HDM.}
	\label{f:diphoton}
\end{figure}

%
We now turn to the constraints arising from measurements of the diphoton signal strength, $\mu_{\gamma\gamma}$, and from direct searches for $H\to\gamma\gamma$. As discussed in Sec.~\ref{s:diphoton}, these observables are primarily sensitive to $f_C$ and $f_H$, respectively. In the left panel of Fig.~\ref{f:diphoton}, we superimpose the region allowed by $\mu_{\gamma\gamma}$ onto the set of parameter points that satisfy the combined constraints from  unitarity, BFB conditions, the $T$-parameter, and direct LHC searches which have been presented in Figs.~\ref{f:Hsearch}, and \ref{f:Csearch}. For further clarity, the scattered background points in the left panel of Fig.~\ref{f:diphoton} correspond to those shown in the first panel of Fig.~\ref{fig:uniMvsf}, with the additional removal of points excluded by the direct search constraints in Figs.~\ref{f:Hsearch}, and \ref{f:Csearch}.
In the left panel of Fig.~\ref{f:diphoton}, the light-pink shaded band with dotted boundary overlaid on the scattered points represents the current $2\sigma$ allowed region from the LHC measurement of $\mu_{\gamma\gamma}$, as given in \Eqn{e:mugam_LHC}. Likewise, the narrower light-purple band with dot-dashed boundary corresponds to the projected $2\sigma$ sensitivity from future measurements at the HL-LHC and the ILC, as summarized in \Eqn{e:kgfuture}. The vertical solid blue line indicates the current constraint from the $b \to s\gamma$ process, which imposes a lower bound on the charged scalar mass, $m_C \gtrsim 580~\mathrm{GeV}$, in the type-II 2HDM.

At first sight, the bounds derived from $\mu_{\gamma\gamma}$ may appear to be overshadowed by the stringent constraint from $b \to s\gamma$. However, it is important to note that the latter is an artefact of our choice of a type-II Yukawa structure. Had we instead adopted a type-I Yukawa structure, no such lower bound on the nonstandard scalar masses from flavor constraints would arise for $\tan\beta \gtrsim 1$~\cite{Das:2015qva}.

By contrast, the constraint $-2.5 \lesssim f_C \lesssim 1.4$ inferred from the left panel of Fig.~\ref{f:diphoton} using the current measurement of $\mu_{\gamma\gamma}$ is independent of the Yukawa sector and is therefore applicable to all 2HDM variants with a softly-broken $Z_2$ symmetry. More importantly, future collider projections for Higgs diphoton measurements are expected to impose significantly stronger bounds, $-1.3 \lesssim f_C \lesssim 0.7$, implying that the scope of contribution of the electroweak VEV to the charged scalar mass will be substantially diminished.

In the right panel of Fig.~\ref{f:diphoton}, we present the constraints in the $f_H$-$m_H$ plane arising from searches for $pp \to H/A \to \gamma\gamma$. The excluded parameter points are shown in red, overlaid on the teal background points. As discussed previously, these background points correspond to the $f_H$-$m_H$ parameter space shown in Fig.~\ref{fig:uniMvsf}, after removing regions already excluded by the searches presented in Figs.~\ref{f:Hsearch} and \ref{f:Csearch}. The current bounds on the $pp \to H/A \to \gamma\gamma$ production cross section~\cite{ATLAS:2021uiz,CMS:2024nht} are sufficiently strong to impose additional constraints on the $f_H$-$m_H$ plane for a type-II 2HDM. We also note that the $pp \to H \to \gamma\gamma$ searches are performed in the mass range $160~\mathrm{GeV} < m_H < 3~\mathrm{TeV}$;~\cite{ATLAS:2021uiz} consequently, parameter points with $m_H <  160~\mathrm{GeV}$ remain unconstrained by $pp \to H \to \gamma\gamma$ searches in the right panel of Fig.~\ref{f:diphoton}. However, there are some red points in $125~\mathrm{GeV} < m_H < 160~\mathrm{GeV}$ range that are disallowed from $pp \to A \to \gamma\gamma$ searches. 

\begin{figure}[htb!]
	\centering
		\includegraphics[scale=0.3]{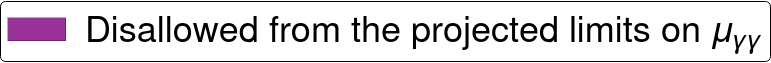}\\
		\vspace{0.5cm}
	\includegraphics[scale=0.3]{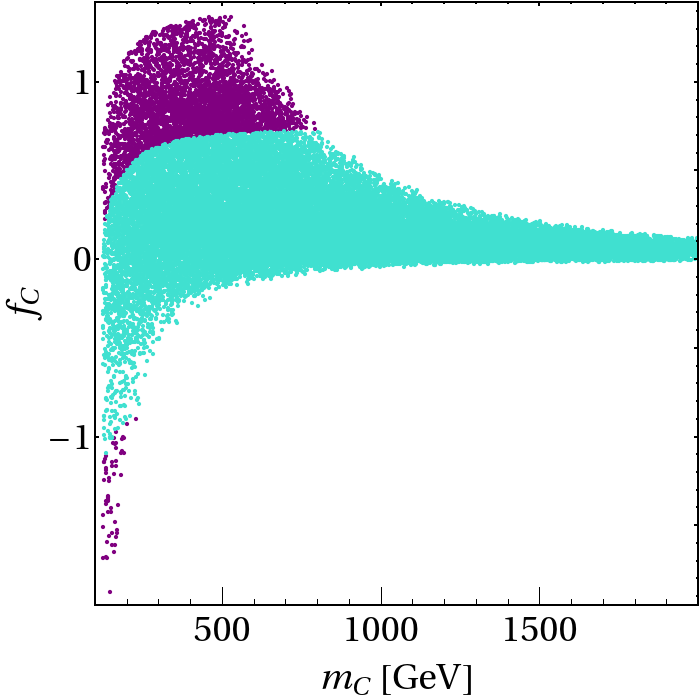}
		\includegraphics[scale=0.3]{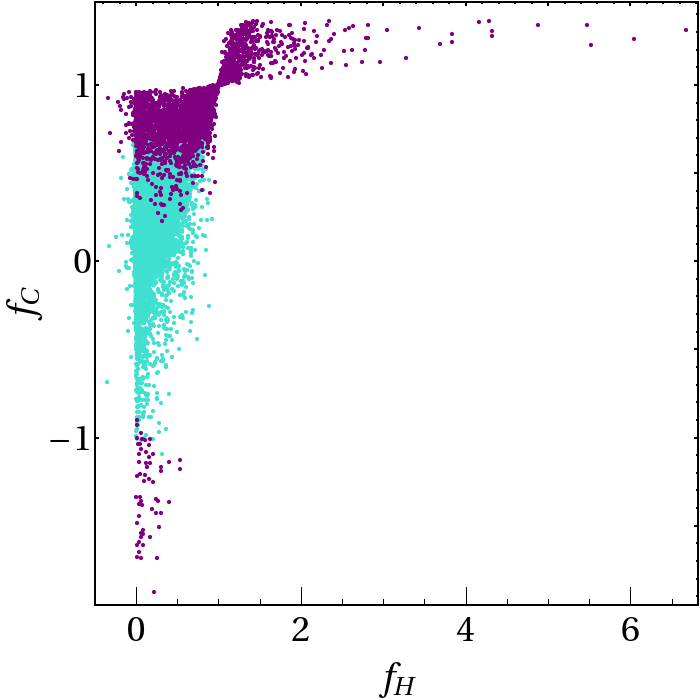}\\
			\vspace{0.5cm}
	\includegraphics[scale=0.3]{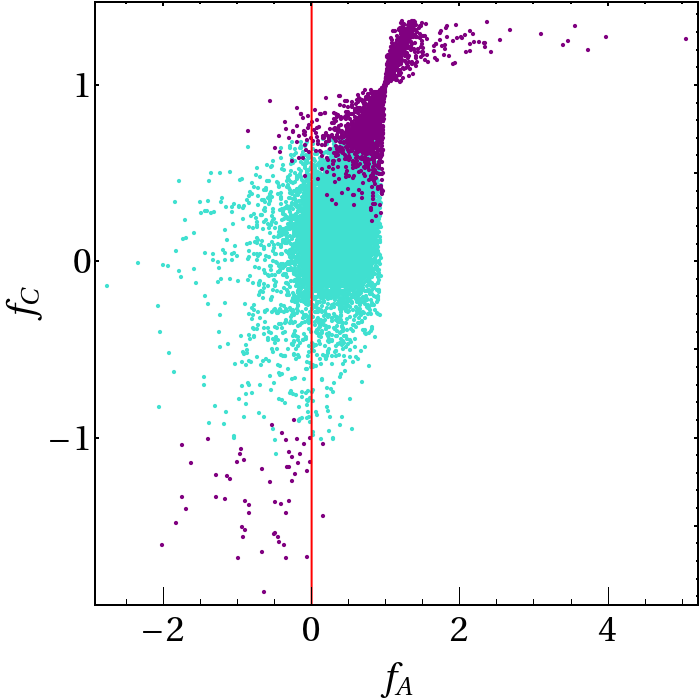}
	\includegraphics[scale=0.3]{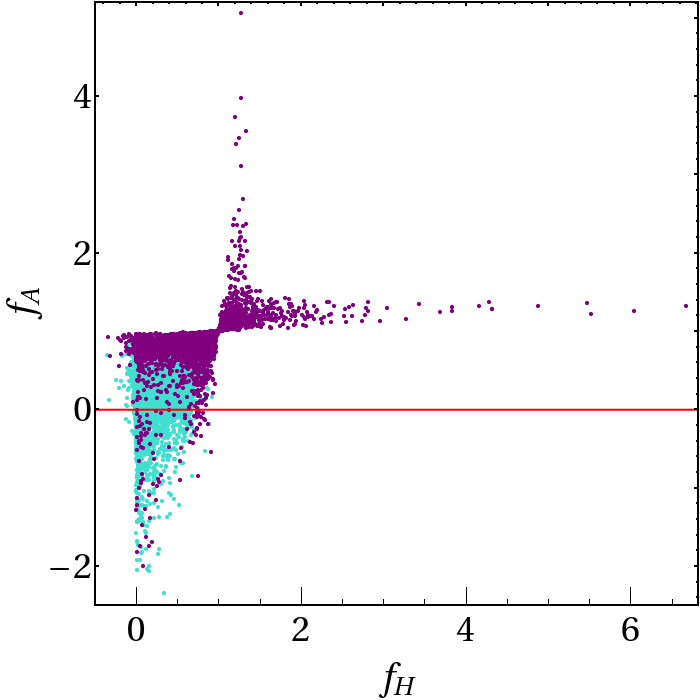}
	\caption{Points satisfying the combined constraints from unitarity, BFB conditions, and $\Gamma_{H,A,C} < m_{H,A,C}$, while remaining consistent with the current measurement of $\mu_{\gamma\gamma}$. Points highlighted in purple are disallowed from the anticipated future measurement of $\mu_{\gamma\gamma}$ with improved precision. These constraints are chosen such that the resulting bounds remain applicable to all variants of the 2HDM with a softly-broken $Z_2$ symmetry. The red solid line at $f_A=0$ corresponds to the case of a 2HDM potential with a softly-broken U(1) symmetry. Additionally, we note that $f_X = 1$ ($X \equiv C, H, A$) implies $\Lambda^2 = 0$ according to \Eqn{e:nonstandrad_frac}, corresponding to the limiting case in which the 2HDM scalar potential possesses an exact $Z_2$ symmetry}
	\label{f:diphoton1}
\end{figure}

Keeping in mind that Fig.~\ref{f:diphoton} is specific to the Type-II Yukawa structure, it would be interesting to investigate how much of a parameter space can be constraint in a model independent manner. To this end, we note that constraints from $\mu_{\gamma\gamma}$ in particular do not depend on the type of the 2HDM in the alignment limit. In view of this, we display in Fig.~\ref{f:diphoton1} all the points that are consistent with the combined constraints from perturbative unitarity, BFB conditions, the $T$-parameter, and $\Gamma_{H,A,C} < m_{H,A,C}$. These points are displayed in the $f_C$-$m_C$ plane and various $f_X$-$f_Y$ ($X,Y \equiv C, H, A$) planes. Consequently, the constraints shown in Fig.~\ref{f:diphoton1} are largely insensitive to the Yukawa structure and are therefore applicable to all four natural flavor-conserving (NFC) realizations of the 2HDM.\footnote{Note that the constraints $\Gamma_{H,A,C} < m_{H,A,C}$ is primarily dictated by the top Yukawa, which remains the same across all the NFC variants of 2HDM.} We wish to reiterate that all the points in Fig.~\ref{f:diphoton1} are consistent with the current experimental determination of $\mu_{\gamma\gamma}$. Additionally, the points highlighted in purple are projected to be excluded at the $95\%$ C.L.\ by future improvements in the high-precision measurements of $\mu_{\gamma\gamma}$.

%



As evident from Fig.~\ref{f:diphoton1}, the fractions $f_{C,H,A}$ are already constrained by current experimental bounds to satisfy $f_{C,H,A} \lesssim \mathcal{O}(1)$. These limits are expected to become progressively more stringent with future high-precision measurements of $\mu_{\gamma\gamma}$, potentially constraining regions with $f_{C,H,A} \approx 1$. This, in view of \Eqn{e:nonstandrad_frac}, implies that a non-negligible fraction of the nonstandard scalar masses must originate from the soft $Z_2$-breaking parameter. This demonstrates the extent to which $\mu_{\gamma\gamma}$ measurements alone can constrain the parameter space that survives theoretical consistency conditions. 



%
 We therefore conclude that the absence of observable new-physics effects in $\mu_{\gamma\gamma}$ and in $pp \to H \to \gamma\gamma$ searches can be interpreted as constraints on the fractions of the electroweak VEV contributing to the nonstandard scalar masses. This interpretation offers an intriguing and complementary perspective to direct collider searches for new physics.

\section{Summary}
\label{s:summary}
The principal motivation of our work is to demonstrate that, within the framework of a 
2HDM, the masses of the nonstandard scalars can be systematically decomposed into two 
distinct contributions: one originating from the electroweak VEV, and the other 
stemming from the soft symmetry-breaking parameter present in the scalar potential. 
This decomposition provides a transparent understanding of the origin of mass scales 
in the extended scalar sector and clarifies the role of soft breaking terms in shaping 
the physical spectrum.

Our analysis begins with the observation that soft-breaking parameters in the 2HDM 
scalar potential are often indispensable for ensuring a consistent {\em decoupling} 
limit~\cite{Gunion:2002zf} of the nonstandard scalars~\cite{Faro:2020qyp,Bhattacharyya:2014oka}. In their absence, achieving heavy nonstandard states 
without jeopardizing perturbative unitarity becomes impossible, especially when the 
additional
symmetry imposed on the scalar potential is spontaneously broken. Motivated by this, 
we first consider the widely studied case of a 2HDM with a softly-broken $Z_2$ 
symmetry, as typically employed in the NFC scenarios. Within this setup, we explicitly 
disentangle the contributions to the nonstandard scalar masses that arise from the 
electroweak VEV from those generated by the soft-breaking parameter, thereby making 
manifest the structural separation between electroweak and non-electroweak mass 
sources.

We then formalize the notion of nonstandard sources of mass by embedding the 2HDM into 
a framework extended by a SM scalar singlet, while maintaining an exact $Z_2$ symmetry 
in the full Lagrangian. In this construction, the effective soft-breaking parameter of 
the low-energy 2HDM emerges dynamically from the VEV of the singlet field. Crucially, 
this singlet VEV is unrelated to the electroweak VEV, thereby providing a genuine 
non-electroweak origin for the soft-breaking term. This observation justifies our 
interpretation of the soft-breaking parameter as a nonstandard mass source for the 
additional scalar states. As a nontrivial consistency check, we explicitly verify 
that, in the alignment limit, the mass of the SM-like Higgs boson is generated 
entirely by the electroweak VEV, reinforcing the separation between standard and 
nonstandard mass-generating mechanisms.

As a complementary illustration, we also argue that a 2HDM potential featuring a 
softly-broken global $U(1)$ symmetry can naturally descend from a high-scale theory 
invariant under a discrete symmetry. In our explicit example, we consider a theory 
endowed with a $Z_3$ symmetry at high energies. The spontaneous breaking of this 
discrete symmetry at a potentially very high scale gives rise to an effective soft 
$U(1)$-breaking parameter in the low-energy theory, again linked to a nonstandard VEV. 
Through these examples, we highlight the distinct UV completion pathways that can lead 
to 2HDM potentials with softly-broken discrete versus continuous Abelian symmetries, thereby 
elucidating their differing structural origins.

To quantify the relative proportion of non-electroweak mass contributions, we 
introduce a convenient parametrization that senses the fraction of the nonstandard 
scalar masses arising from nonstandard sources. We emphasize that these fractions are 
subject to both theoretical and phenomenological constraints. On theoretical grounds, 
we demonstrate that if the nonstandard scalar masses are required to lie significantly 
above the electroweak scale, perturbative unitarity constraints enforce that the 
dominant portion of these masses must originate from non-electroweak sources.

On the experimental side, we examine the constraining power of Higgs signal strength 
measurements and direct searches. In particular, we highlight the role of the diphoton signal strength modifier $\mu_{\gamma\gamma}$ and searches for $pp \to H \to 
\gamma\gamma$ in constraining the mass fractions $f_C$ and $f_H$, respectively. We 
show that current measurements of $\mu_{\gamma\gamma}$ already impose nontrivial 
bounds on $f_C$, and that the projected precision at the HL-LHC can further tighten 
these constraints to the range $-0.8 < f_C < 1$. Moreover, in the custodial limit of 
the 2HDM, characterized by $m_C = m_A$, the bounds on $f_C$ can be directly translated 
into corresponding constraints on $f_A$. Collectively, these results demonstrate that 
both theoretical consistency and present and future collider data significantly 
restrict the possible electroweak origin of nonstandard scalar masses in the 2HDM 
framework.

Thus, we may conclude that the seemingly null results in BSM searches within the diphoton 
channels should not be dismissed as mere non-findings. Instead, these results are directly sensitive to the proportion of the standard electroweak VEV contributing to the masses of nonstandard particles. Consequently, they offer valuable insights into the underlying composition of these nonstandard mass terms, highlighting the critical importance of continuing such experimental investigations.

\paragraph{Acknowledgements:}
DD thanks the ANRF (Erstwhile SERB), India for financial support
through grant no. CRG/2022/000565.
IS acknowledges the financial support
through grant no. MTR/00715/2023 from SERB MATRICS project by ANRF (Erstwhile SERB), India.
SP acknowledges the Department of Science and Technology~(DST)-INSPIRE, Government of India, for the fellowship awarded under sanction no. DST/INSPIRE Fellowship/2021/IF210680.
AS thanks Anusandhan National Research Foundation (ANRF) for providing the necessary financial support through the SERB-NPDF grant
(Ref. No.: PDF/2023/002572) and also acknowledges the support from the Department of Atomic Energy (DAE), India, for the Regional Centre for Accelerator-based Particle Physics (RECAPP), Harish Chandra Research Institute.
\appendix
\setcounter{equation}{0}
\renewcommand{\theequation}{\thesection.\arabic{equation}}
\section{General Expressions for the trilinear couplings}
\label{app:gen_coup}
For a 2HDM potential with a softly-broken $Z_2$ symmetry, the general expressions for the self-couplings appearing in \Eqn{e:triself} are given below:
\begin{subequations}
	\label{e:triselfgen}
	\begin{eqnarray}
		\lambda_{hH^+H^-} &=& \frac{1}{2v\sin 2\beta}\left[ (2m_C^2-m_h^2)
		\sin(\alpha-3\beta) +(4\Lambda^2-2m_C^2-3m_h^2)\sin(\alpha+\beta) \right] 
		\,, \\
		\lambda_{HH^+H^-} &=& \frac{1}{2v\sin 2\beta}\left[ (2m_C^2-m_H^2)
		\cos(\alpha-3\beta) +(4\Lambda^2-2m_C^2-3m_H^2)\cos(\alpha+\beta) \right] 
		\,, \\
		\lambda_{hHH} &=& \frac{\cos(\alpha-\beta)}{v\sin 2\beta}\left[
		(3\Lambda^2-2m_H^2-m_h^2)\sin 2\alpha-\Lambda^2\sin2\beta \right] 
		\,, \\	
		\lambda_{hAA} &=& \frac{1}{2v\sin 2\beta}\left[ (2m_A^2-m_h^2)
		\sin(\alpha-3\beta) +(4\Lambda^2-2m_A^2-3m_h^2)\sin(\alpha+\beta) \right] 
		\,, \\
		\lambda_{HAA} &=& \frac{1}{2v\sin 2\beta}\left[ (2m_A^2-m_H^2)
		\cos(\alpha-3\beta) +(4\Lambda^2-2m_A^2-3m_H^2)\cos(\alpha+\beta) \right] 
		\,, \\   
		\lambda_{HHH} &=& \frac{1}{2v\sin 2\beta}\Big[ \Lambda^2
		\cos(\alpha-3\beta) +(\Lambda^2-m_H^2)\cos(3\alpha-\beta) \nonumber \\ && +(2\Lambda^2-3m_H^2)\cos(\alpha+\beta) \Big] 
		\,.
	\end{eqnarray}
\end{subequations}

\section{Electroweak $T$-parameter in nHDMs with scalar singlets}\label{sec:custodial}

In this appendix, we provide the conditions for the $T$-parameter to vanish in models with an arbitrary number of SM doublets (nHDMs) extended by real and complex SM singlets ($Y=0$).  
The goal here is to recast the results of Ref.~\cite{Grimus:2007if} for the case where all singlets are neutrally-charged, in a notation which can be easily translated to the cases at hand in the main text.  

We start by defining the fields, after SSB, as 
\begin{subequations}
	\begin{eqnarray}
		k=1, \dots, n_d \, : \quad && \phi_k = \begin{pmatrix} \omega^+_k \\ \varphi_k^0 \end{pmatrix} = \begin{pmatrix} \omega^+_k \\ (v_k + h_k +i \, z_k)/\sqrt{2} \end{pmatrix} \, , \\
		j=1, \dots, n_c \, : \quad &&  \chi_j = (u_j + \chi^0_j +i \, \xi_j)/\sqrt{2}  \, ,  \\
		l=1, \dots, n_r \, : \quad && \rho_l = u_l + \rho^0_l   \, ,  
	\end{eqnarray}
\end{subequations}
where $n_d$ is the number of $SU(2)$ doublets, $n_c$ is the number of $SU(2)$ complex singlets, and $n_r$ counts the $SU(2)$ real singlets, and the quantities $v_k$, $u_j$ and $u_l$ denote VEVs of the respective fields.  

Under the assumption that the electrically neutral fields can be classified as CP-even and CP-odd, we define the mass matrices as
\begin{subequations}
	\begin{eqnarray}
		V_{\rm mass}^{\rm charged} &=& \omega^-  \, \mathcal{M}_C^2 \,\,  \omega^+ \, , \\
		V_{\rm mass}^{\textrm{ CP-odd}} &=& \begin{pmatrix} z_k & \xi_j \end{pmatrix} \frac{\mathcal{M}_P^2}{2} \begin{pmatrix} z_k \\ \xi_j \end{pmatrix} \, , \\
		V_{\rm mass}^{\textrm{ CP-even}} &=& \begin{pmatrix} h_k   & \chi_j^0   & \rho_l^0 \end{pmatrix} \frac{\mathcal{M}_H^2}{2} \begin{pmatrix} z_k \\ \chi_j^0 \\ \rho_l^0 \end{pmatrix} \, , 
	\end{eqnarray}
\end{subequations}
where $\mathcal{M}_C^2$ is the $n_d \times n_d$ mass-squared matrix in the charged 
scalar sector. Similarly, $\mathcal{M}_P^2$ is the $n_P \times n_P$ ($n_P = n_d + n_c$) mass-squared matrix in the CP-odd 
scalar sector, and $\mathcal{M}_H^2$ is the $n_H \times n_H$ ($n_H = n_d + n_c + n_r$) mass-squared matrix in the CP-even 
scalar sector.  
Finally, we also define the matrices that rotate the fields in the Lagrangian basis into the fields in the mass basis:
\begin{subequations}
	\begin{eqnarray}
		D_C^2 &=& \mathcal{R}_C^T \cdot \mathcal{M}_C^2 \cdot \mathcal{R}_C \, , \\
		D_P^2 &=& \mathcal{R}_P^T \cdot \mathcal{M}_P^2 \cdot \mathcal{R}_P \, , \\
		D_C^2 &=& \mathcal{R}_H^T \cdot \mathcal{M}_H^2 \cdot \mathcal{R}_H \, , 
	\end{eqnarray}
\end{subequations}
where, the matrices $D_X^2$ ($X\equiv C,P,H$) are the diagonal mass-squared matrices, ordered such that $(D_X^2)_{ii} = \mu^X_i$, with the understanding $\mu_1^C=\mu_1^A=0$ (the Goldstones) and $\mu_1^H=m_h^2$ (the 125 GeV scalar).  
At this point, we note that these (real orthogonal) rotation matrices have different dimensions ($n_d$ for $\mathcal{R}_C$, $n_P$ for $\mathcal{R}_P$, and $n_H$ for $\mathcal{R}_H$), but the contributions for the $T$-parameter come solely from the $SU(2)$ doublets.  
As such, it is useful to break down these matrices as 
\begin{eqnarray}
	\mathcal{R}_H = \begin{pmatrix} \left(\mathcal{R}_H^{n_d}\right)_{n_d \times n_H} \\ \left(V_H\right)_{(n_H-n_d) \times n_H} \end{pmatrix} \, , \qquad \mathcal{R}_P = \begin{pmatrix} \left(\mathcal{R}_P^{n_d}\right)_{n_d \times n_P} \\ \left(V_P\right)_{(n_P-n_d) \times n_P} \end{pmatrix} \, .
\end{eqnarray}
The objective behind this segregation may be understood from the fact that the elements of $V_P$ and $V_H$ do not enter the expression of the $T$-parameter. 
Using these definitions, we can write the general expression of $\Delta T_\text{NP}$ as~\cite{Grimus:2007if}
\begin{subequations}\label{eq:allparts}
	\begin{eqnarray}
		16 \pi^2 \sin^2\theta_w M_W^2 \Delta T_\text{NP} &=&  \sum_{a=2}^{n_d} \sum_{b=2}^{n_P} \left(\mathcal{R}_C^T \, \mathcal{R}_P^{n_d} \right)_{ab}^2 F(\mu^C_a \, , \mu^P_b)  \label{eq:part1}\\
		&&+  \sum_{a=2}^{n_d} \sum_{b=1}^{n_H} \left[\mathcal{R}_C^T \, \mathcal{R}_H^{n_d} \right]_{ab}^2 F(\mu^C_a \, , \mu^H_b)  \label{eq:part2}  \\
		&&-  \sum_{a=2}^{n_d} \sum_{b=1}^{n_H} \left[(\mathcal{R}_P^{n_d})^T \, \mathcal{R}_H^{n_d} \right]_{ab}^2 F(\mu^P_a \, , \mu^H_b)  \label{eq:part3}  \\
		&&+  \sum_{a=2}^{n_c} \sum_{b=1}^{n_H} \left[(\mathcal{R}_P^{n_d})^T \, \mathcal{R}_H^{n_d} \right]_{a+n_d, b}^2 F(\mu^P_{a+n_d} \, , \mu^H_b)   \label{eq:part4}  \\
		&&+  3 \sum_{b=1}^{n_H}  \left[(\mathcal{R}_P^{n_d})^T \, \mathcal{R}_H^{n_d} \right]_{1 b}^2  \left\{ F(M^2_Z \, , \mu^H_b) -  F(M^2_W \, , \mu^H_b) \right\}  \label{eq:part5} \\
		&&- 3 \left\{ F(m^2_Z \, , m^2_h) -  F(m^2_W \, ,m^2_h) \right\}    \label{eq:part6}\, ,
	\end{eqnarray}
\end{subequations}
where $F(x,y)$ is already defined in Eq.~\eqref{eq:Ffunc}.

From the results of Refs.~\cite{Kundu:2021pcg, Das:2022gbm}, we expect that the charged-scalars from the $SU(2)$ doublets should be placed in custodial $SU(2)_C$ triplets along with the pseudoscalars that arise purely from the mixing among the components of the $SU(2)$ doublets.  
For this to happen, we require~\cite{Das:2022gbm}  
This can be achieved by imposing
\begin{eqnarray}\label{eq:cond1}
	\left( \mathcal{R}_C^T \mathcal{R}_P^{n_d} \right)_{ab} = \delta_{ab} \, , \qquad \mu^C_a = \mu^P_a \, , \quad a=1, \dots, n_d \, . 
\end{eqnarray}
To understand this condition, it is instructive to decompose $\mathcal{R}^{n_d}$ as 
\begin{eqnarray}
	\mathcal{R}^{n_d} = \begin{pmatrix} \left( \mathcal{O}^{n_d}_P \right)_{n_d \times n_d} \, , & \left( \mathcal{O}^{n_c}_P \right)_{n_d \times n_c}
	\end{pmatrix} \, .
\end{eqnarray}
Keeping in mind the fact that $\mathcal{R}_P^T\mathcal{R}_P=\mathbb{1}$, the first condition in Eq.~\eqref{eq:cond1} reduces to
\begin{eqnarray} \label{eq:cond2}
	\mathcal{O}_P^{n_d} = \mathcal{R}^C \, , \quad \text{and} \quad \mathcal{O}^{n_c}_P = 0 \, .
\end{eqnarray}
The meaning of this is rather simple: the masses and mixings of the pseudoscalars arising from the $SU(2)$ doublets must be identical to those of the charged-scalars, and the pseudoscalars from the $SU(2)$ singlets cannot mix with these states (but are otherwise unrestricted).  
Reverting to the custodial multiplets interpretation, the pseudoscalars and charged-scalars arising from $\phi_k$ will be in triplets, whereas the pseudoscalars of $\chi_j$ will be custodial singlets.  
Lastly, since the CP-even states are all custodial singlets~\cite{Das:2022gbm,Nishi:2011gc}, there are no restrictions on the masses and mixings in the CP-even sector.  
\begin{itemize}
	\item \Eqn{eq:part1} is identically zero, since $F(x, x)=0$ ,
	\item \Eqn{eq:part2} cancels exactly with \Eqn{eq:part3} ,
	\item \Eqn{eq:part4} is zero, since $\mathcal{O}^{n_c}_P = 0$ .
\end{itemize}

Finally, it is easy to see that the contributions from \Eqs{eq:part5}{eq:part6} cancel each-other in the alignment limit.  
To verify this, we start by defining a convenient Higgs basis in our case by the transformation of $\phi_k$ that diagonalizes the charged-scalar mass matrix, while not transforming the singlets (similar to the charged-Higgs basis of Ref.~\cite{Bento:2017eti}).  
Then, we can express this transformation of the neutral fields ($\mathcal{R}_\beta$) and the alignment condition as:
\begin{eqnarray}\label{eq:alignmentCondition}
	\mathcal{R}_\beta = \begin{pmatrix} \left(\mathcal{R}_C\right)_{n_d \times n_d} & 0 \\ 0 &  \mathbb{1}_{(n_H-n_d) \times (n_H-n_d)} \end{pmatrix} \, , \quad \left( \mathcal{R}_\beta^T \mathcal{R}_H \right)_{11} =\left( \mathcal{R}_C^T \mathcal{R}^{n_d}_H \right)_{11} = 1 \, .
\end{eqnarray}
To calculate the quantity in \Eqn{eq:part5}, we now note 
\begin{eqnarray}
	&& 3 \sum_{b=1}^{n_H}  \left[(\mathcal{R}_P^{n_d})^T \, \mathcal{R}_H^{n_d} \right]_{1 b}^2  \left\{ F(M^2_Z \, , \mu^H_b) -  F(M^2_W \, , \mu^H_b) \right\}  \nonumber \\
	&=& 3 \left[(\mathcal{R}_P^{n_d})^T \, \mathcal{R}_H^{n_d} \right]_{1 1}^2  \left\{ F(M^2_Z \, , m^2_h) -  F(M^2_W \, , m^2_h) \right\} + (\dots) \, , 
\end{eqnarray}
which, after imposing the conditions in \Eqs{eq:cond1}{eq:alignmentCondition} exactly cancels the expression given in \Eqn{eq:part6}. 

Thus, to summarize, we can have $\Delta T_\text{NP}=0$ by design, if we impose the following:
\begin{itemize}
	\item The alignment limit:
	\begin{eqnarray}
		\left( \mathcal{R}_C^T \mathcal{R}_H^{n_d}\right)_{11} = 1  \, ,
	\end{eqnarray}
	which, in the conventions of the main text becomes (c.f. with \Eqs{e:alignment1_2hdms}{e:alignment1_2hdms_u1})
	\begin{eqnarray}
		\left( \mathcal{O}_\beta \mathcal{O}_\alpha^T \right)_{11} = 1\Rightarrow \cos\left( \alpha -\beta\right)=0 \quad \text{and } \quad \epsilon =0 \, .
	\end{eqnarray}
	\item Zero mixing between the pseudoscalars from the $SU(2)$ doublets and from the $SU(2)$ singlets:
	\begin{eqnarray}
		\mathcal{R}_P  = \begin{pmatrix} \mathcal{O}_P^{n_d} & 0 \\ 0 & \mathcal{O}'_P  \end{pmatrix}  \, ,
	\end{eqnarray}
	which is automatically satisfied for the case of Sec.~\ref{sec:Z2-2HDMS}. 
	However, for the case of Sec.~\ref{sec:U1-2HDM}, we additionally require (see \Eqn{e:Odelta})
	\begin{eqnarray}
		\delta=0 \, .
	\end{eqnarray}
	\item A pseudoscalar arising purely from the $SU(2)$ doublets should have the same mass as the corresponding charged-scalar, and the two sectors should have the same mixings:
	\begin{eqnarray}
		\mathcal{R}_C = \mathcal{R}_P^{n_d} \quad \text{and} \quad m^C_a = m^P_a\, , \quad a, =1, \dots, n_d \, .
	\end{eqnarray}
\end{itemize}

\bibliographystyle{JHEP}
\bibliography{Soft-Origin}

\end{document}